\documentclass[%
aps,
 reprint,
 amsmath,amssymb,
prper,
floatfix
]{revtex4-2}

\usepackage{graphicx}
\usepackage{dcolumn}
\usepackage{bm}
\usepackage{tikz}
\usepackage{pgfplots}
\usepackage[export]{adjustbox}
\pgfplotsset{compat=1.17}
\usepackage{cleveref}
\usepackage{rotating}
\usepackage{enumitem}

\begin{document}


\title{Assessing Confidence in AI-Assisted Grading of Physics Exams through Psychometrics: An Exploratory Study}

\author{Gerd Kortemeyer}
 \email{kgerd@ethz.ch}
 \affiliation{%
Rectorate and AI Center, ETH Zurich, 8092 Zurich, Switzerland
}%
\altaffiliation[also at ]{Michigan State University, East Lansing, MI 48823, USA}

\author{Julian N{\"o}hl}
 \email{jnoehl@ethz.ch}
 \affiliation{%
Institute of Energy \& Process Engineering, ETH Zurich, Tannenstrasse 3, 8092 Zurich, Switzerland
}%

\date{\today}

\begin{abstract}
This study explores the use of artificial intelligence in grading high-stakes physics exams, emphasizing the application of psychometric methods, particularly Item Response Theory (IRT), to evaluate the reliability of AI-assisted grading. We examine how grading rubrics can be iteratively refined and how threshold parameters can determine when AI-generated grades are reliable versus when human intervention is necessary. By adjusting thresholds for correctness measures and uncertainty, AI can grade with high precision, significantly reducing grading workloads while maintaining accuracy. Our findings show that AI can achieve a coefficient of determination of $R^2 \approx 0.91$ when handling half of the grading load, and $R^2 \approx 0.96$ for one-fifth of the load. These results demonstrate AI's potential to assist in grading large-scale assessments, reducing both human effort and associated costs. However, the study underscores the importance of human oversight in cases of uncertainty or complex problem-solving, ensuring the integrity of the grading process. 
\end{abstract}

\maketitle

\section{Introduction}
\subsection{AI-supported Grading}
The rapid development of artificial intelligence (AI) is opening new opportunities  in many scientific fields. Most recently, Large Language Models (LLMs), in particular generative Pre-trained Transformers (GPT)~\cite{chatgpt}, such as GPT-4~\cite{gpt4}, have demonstrated impressive abilities in solving and evaluating academic tasks~\cite{meyer2023chatgpt},  including tasks related to academic education~\cite{kasneci2023chatgpt}. Consequently, AI is increasingly integrated into physics education~\cite{yeadon2024impact,sperling2024artificial,polverini2024understanding}, including aiding educational research~\cite{tschisgale2023integrating}, solving problems~\cite{kung2022,achiam2023gpt,kortemeyer23ai,lopez2024challenging,wang2024examining}, constructing new problems~\cite{kuchemann23}, and promising preliminary results for grading problem solutions~\cite{wilson22,wan24,kortemeyer24aigrading,kortemeyer2024grading,liu2024ai}.

However, the reliability of AI in grading complex problem-solving processes is still a concern~\cite{ding2023students,dahlkemper23,li2023wrong}; risks associated with inadequate determination of confidence levels have garnered immediate attention, so, for example, within the European Union, ``AI systems intended to be used to evaluate learning outcome'' are considered ``high-risk''~\cite{euactannex}, and as such, human oversight is obligatory~\cite{euactobl}\footnote{Switzerland is not part of the European Union, but often aligns itself with European regulations due to its close economic and political ties with EU member states.}. This study examines the use of psychometric methods, such as Item Response Theory (IRT), to determine when AI grading can be trusted and when human intervention is necessary. In a ``human-in-the-loop'' workflow~\cite{de2020case}, AI assists in grading, but humans always retain the final decision, especially in cases where diverse solution paths or handwriting complexities challenge automated systems~\cite{kashyd01,alsalmani23}. We focus on understanding where AI can be relied upon and where human oversight is essential, aiming to use psychometrics to guide this balance.

\subsection{Psychometrics}
Psychometrics, as a scientific discipline, has evolved substantially since its origins in the early 20th century, where it sought to develop standardized methods for measuring psychological traits like intelligence and aptitude with an unfortunate link to eugenics~\cite{spearman1914heredity}. It introduced the concepts of factor analysis, which identify underlying relationships between measured variables, aiming to reduce them into a smaller set of factors that represent common traits or dimensions within the data. Early on, one such factor was the general intelligence factor, $g$, which was used to quantify human abilities~\cite{hart1912general,thurstone1934vectors,spearman1961general}. As the field matured, broad-stroke measures like ``general intelligence'' were replaced by measures of domain-specific abilities (which grow through learning), and psychometrics were increasingly applied to the assessment instruments rather than the test-takers, where they became a cornerstone for ensuring the validity, reliability, and fairness of tests.

Validity refers to how well a test measures what it claims to measure. For example, a question might be too easy or too difficult to provide meaningful assessment, or it might include misleading statements or bias that lets test takers answer it correctly or incorrectly independent of their mastery of the subject that is supposed to be tested. Psychometric methods help establish validity through the lenses of both content validity and construct validity. Content validity ensures that items represent the domain they are supposed to assess, while construct validity examines whether test items truly measure the underlying theoretical construct, such as ``ability'' in a given subject. Both forms of validity are key when designing tests and when evaluating the appropriateness of items, that is, they judge the test items under the assumption that their grading is carried out without errors and perfect.

In this study, we assume that the test items have content and construct validity, but that the AI-based grading is error-prone and imperfect. We thus add another lens to psychometrics: grading validity, which encompasses
\begin{itemize}
\item the rules for grading, in our case the prompts for the AI~system, and
\item the grading judgement, in our case the inference and reasoning performance of the AI~system.
\end{itemize}

As a point of nomenclature: within psychometrics, what physicists would call a ``problem'' or ``test question'' is usually referred to as an ``item.'' Since our exam is graded on a rubric, we apply the framework to rubric items; a particular test taker may or may not receive credit on a particular rubric item.

\subsection{Classical Test Theory}
Classical Test Theory (CTT) primarily focuses on raw scores and descriptive statistics. CTT is both reliable and useful for assessing overall test performance by summarizing test scores with metrics like mean, variance, and standard deviation. Its strength lies in its simplicity and ease of implementation, making it widely used in educational assessments and standardized testing. We are going to be using some of these techniques when refining the grading rules as part of establishing grading validity.

CTT assumes that each test-taker has a true score that reflects their ability, with observed scores deviating from this true score due to random errors~\cite{devellis2006classical}. This framework works well when the goal is to evaluate the overall difficulty of a test or to compare test-takers across different tests. However, CTT has limitations when it comes to predicting how a particular test-taker will perform on individual items. Since it relies solely on aggregate statistics like total scores, it cannot account for how an individual's ability interacts with the difficulty and discrimination of specific items. As a result, CTT provides no direct mechanism for modeling item-level performance or adapting assessments based on a test-taker's latent traits.

This is where methods like Item Response Theory offer useful approaches, many of which are rooted in Bayesian statistics. While CTT treats test scores as static outcomes, Bayesian models allow for probabilistic predictions~\cite{hambleton1993comparison}. Bayesian statistics can model the probability of a correct response on each item based on the test-taker's latent ability and the item's properties.

\subsection{Item Response Theory}

Item Response Theory (IRT) offers the opportunity for data-driven development and evaluation of assessment items, such as homework, practice, concept inventory, and exam problems.  In essence, it predicts the probability that a particular test taker gets a particular test item correct. The technique is increasingly used in Physics Education Research, for example to examine the validity of concept tests~\cite{lin09,cardamone11}, but also to evaluate online homework problems (e.g.~\cite{lee08,kortemeyer2014b}).

Unlike CTT, IRT assumes that learners possess a latent trait called ``ability'' that goes beyond their observed scores on a particular set of test items. Ideally, test items distinguish (``discriminate'') between learners who have high ability and those who have low ability: high-ability students should be likely to solve the item correctly, low-ability students should not. However, in reality,  a learner's ability may result in different scores on different test items, depending on how difficult, well-written, meaningful, or representative these items are. Even high-ability students are likely to fail on an overly difficult item; such an item is not very meaningful on a test, because it is too hard. Even more concerning, high-ability students might be less likely than low-ability students to succeed on poorly written items that contain subtle distractions, missing assumptions, or undesired complications that lower-ability students may not even notice; such an item is detrimental to the meaningfulness of the overt score on the test, which, after all, determines the grade (the latent ability is not meant to determine grades, it is simply a parameter in a Bayesian probability model~\cite{hambleton1993comparison}, though, for a well-written test, the two are correlated~\cite{aka2020}).
\begin{figure}
\begin{center}
\includegraphics[width=\columnwidth]{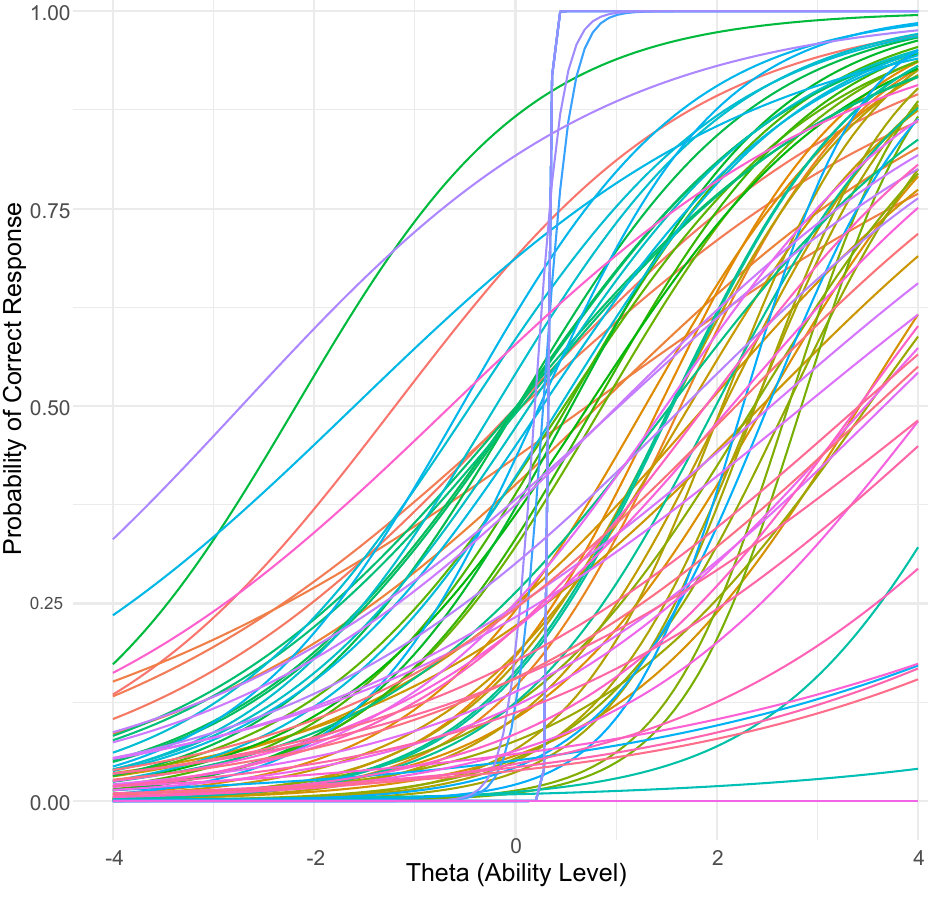}
\end{center}
\caption{Item Characteristic Curves of the rubric items in Sect.~\ref{sec:rubric}, discussed in Sect.~\ref{sec:refine}.}
\label{fig:irtcurves}
\end{figure} 

IRT models the probability $p_i(\theta_j)$ that a learner $j$ with latent ability $\theta_j$ will successfully solve item $i$. As a function of ability, the graph of $p_i(\theta_j)$ for good items is expected to exhibit a stretched S-shape: low-ability students have a low chance of getting the item correct, while high-ability students have a high chance, see Fig.~\ref{fig:irtcurves} as an example. Different IRT models vary in how this so-called ``Item Characteristic Curve'' (ICC) is parameterized, most notably in how many parameters are used to fit the data. This study uses the widely accepted two-parameter logistic (2PL) model~\cite{birnbaum}, which includes a difficulty parameter, $b_i$, and a discrimination parameter, $a_i$. The difficulty parameter, $b_i$, shifts the point of inflection along the ability axis, meaning that higher values of $b_i$ indicate items that require a higher ability level to be likely answered  correctly. Meanwhile, the discrimination parameter, $a_i$, determines the slope of the ICC: higher $a_i$ values result in a steeper slope, indicating that the probability of a correct answer increases more sharply as ability $\theta_j$ rises. The probability in this model is:
\begin{equation}\label{eq:2PL}
p_i(\theta_j) = \frac{1}{1 + \exp\left(-a_i (\theta_j - b_i)\right)}
\end{equation}
While this parametrization may seem arbitrary (``why would the performance follow such a function?'')  and phenomenological, in actual exam scenarios, it has been shown to fit the data rather well~\cite{kortemeyer2019quick}.
Determining the individual learner abilities $\theta_j$ and the item parameters $a_i$ and $b_i$ is an iterative process that minimizes the discrepancy between the predicted probabilities $p_i(\theta_j)$ and the actual performance of learner $j$ on item $i$~\cite{bergner2012model}. This is an optimization process of maximizing the likelihood of the ability estimates. Starting with initial values, the vectors $\theta_j$, $a_i$, and $b_i$ ideally converge to achieve this maximum likelihood estimate; however, convergence is not guaranteed.

\subsection{Research Questions and Hypotheses}
In our explorations, we take content and construct validity as a given, that is, we assume that the high-stakes exam we are using as a dataset has been subject to extensive scrutiny based on years of experience by the exam authors. We are trying to mimic the human grading process by not considering the ground truth until the final evaluation step of our study.

\subsubsection{Rubric Refinement}
We attempt to mimic the iterative refinement of the grading rubric which frequently occurs in human grading of exams as flaws and shortcomings are discovered while applying the grading rules to the students' solutions. We are now asking the following question:
\begin{quote}
Can classical and item-response psychometrical measures be used to assess the grading validity of an AI-based process, both in terms of grading rules (prompts) and grading judgement (inference and reasoning performance)?
\end{quote}
Due to the nature of our dataset, the latter includes the AI system's ability to recognize and make sense of handwritten mathematical derivations and graphical representations.

In particular, we focus on two steps of exam grading that are typically carried out solely by humans:
\begin{itemize}
\item Iterative refinement of the grading rubric
\item Assignment of points
\end{itemize}
In human-grading practice, these two tasks are frequently interwoven: grading may start according to a pre-set rubric, but then it turns out that segments of the examinee population solve problems in unexpected ways, frequently make the same unanticipated error, or appear to consistently misunderstand certain task description.Typically, the grading rubric is adjusted, which unfortunately may also necessitate re-grading already completed examinee solutions according to the new rubric. Our  first hypothesis is:
\begin{quote}
Not referring to specific answers or the ground truth, psychometric methods can be used to iteratively improve the grading rubric.
\end{quote}
\subsubsection{Grading Acceptance}
It is neither likely nor desirable that high-stake exams get solely graded by AI; instead, the AI can assist in grading. In this scenario, it is crucial to assess the validity of AI-grading results on a per-student-per-item base and automatically decide whether to accept the AI-grading result or reject it and have the rubric item for that learner assessed by human. Here, the predictive nature of Item Response allows us to define a measure of uncertainty in the grading as the difference between expected partial credit and the partial credit assigned by the AI. Overall, we consider three parameters:
\begin{description}
\item[Correctness threshold] the minimum partial credit given by the AI above which the item is considered correct. For example, if the AI assigns 30\% partial credit, but the correctness threshold is $0.5$, no credit for the rubric item would be given; if, on the other hand, the correctness threshold is $0.1$, credit would be given. This threshold will be denoted as $C$.
\item[Uncertainty threshold] the maximum uncertainty as determined by Bayesian statistics below which the AI grading is considered trustworthy. For example, if the uncertainty is $0.4$, but the uncertainty threshold is $0.2$, the AI-based rubric-item grading would be discarded, and TAs would need to grade it. This threshold will be denoted as $U$.
\item[Considering only items graded as correct] a boolean parameter that items which the AI grades as incorrect will automatically be reviewed by humans, regardless of uncertainty. With this parameter in effect, only TAs would have the authority to grade a rubric item as incorrect. This parameter will be denoted as $P$.
\end{description}
Our second hypothesis is:
\begin{quote}
Not referring to specific answers or the ground truth, the predictive power of Bayesian models can be used to assess the validity of AI-grading judgements on a per-student-per-item base.
\end{quote}
Based on these confidence measures, it can be decided which items for which students need to be graded by humans.

\section{Setting}
\subsection{Institution}
ETH Zurich is a technical university with around 25,000 students from 120 countries, about one-third of whom identify as female. Admission is highly selective for international students, while Swiss high school graduates have unrestricted access. Most undergraduate courses are conducted in German. Following the academic tradition of German-speaking universities, ETH Zurich focuses on summative assessments at the end of courses, rather than continuous assessments throughout the semester. 
\subsection{Exam}
As dataset, we used a high-stakes exam for engineers on thermodynamics~\cite{kortemeyer2024grading}, dealing with standard topics of energy, exergy, entropy, and enthalpy. Students needed to provide handwritten solutions including derivations, using permanent pens and scribbling out anything they did not want graded.
\begin{itemize}
\item Problem~1 deals with a reactor in steady-state operation involving liquid inflow and outflow, a chemical reaction with heat generation, and a cooling jacket. The students had to:
    \begin{enumerate}[label=\alph*)]
        \item calculate the heat transfer to the cooling fluid;
        \item determine the thermodynamic mean temperature of the cooling fluid;
        \item calculate the entropy production due to heat transfer;
        \item calculate the required mass of water for cooling the reactor from its operating to a lower temperature using an energy balance after stopping the outlet flow;
        \item determine the change in entropy of the reactor contents between the initial and cooled states.
   \end{enumerate}

\item Problem~2 investigates the operation of an aircraft engine consisting of compressors, turbines, nozzles, and heat addition. The students were expected to:
   \begin{enumerate}[label=\alph*)]
        \item draw the engine process qualitatively in a $T-s$ diagram, marking relevant states;
        \item determine the exit speed and temperature of the aircraft engine;
        \item calculate the specific exergy increase between two states;
        \item calculate the specific exergy loss related to the mass flow rate.
   \end{enumerate}

\item Problem~3 involves a hot gas and a solid-liquid system in an isolated cylinder separated by a heat-transferring membrane. The students had to:
    \begin{enumerate}[label=\alph*)]
        \item calculate the initial pressure and mass of the gas in the cylinder;
        \item determine the temperature and pressure of the gas after reaching equilibrium through heat transfer;
        \item calculate the heat transferred from the gas to the ice-water mixture;
        \item calculate the ice content in the second state.
   \end{enumerate}

\item Problem~4 describes a two-step freeze-drying process for food preservation. The students had to:
    \begin{enumerate}[label=\alph*)]
        \item draw the freeze-drying process in a $p$-$T$ diagram, including labeled phase regions;
        \item determine the required mass flow rate of the refrigerant R~134a;
        \item determine the vapor fraction of the refrigerant immediately after throttling;
        \item calculate the coefficient of performance for the cooling cycle;
        \item discuss how the temperature inside the freeze-dryer changes if the cooling cycle continues unchanged.
   \end{enumerate}
\end{itemize}

In the German university tradition, as seen at ETH Zurich, exams often essentially follow a pass/fail structure. Due to the high-stakes nature of these exams, each problem is graded by two teaching assistants (TAs), using a detailed rubric that awards points for individual solution steps. The exam tasks are designed to assess a basic understanding of Thermodynamics, and the course personnel set specific sub-point thresholds for each task to determine whether a student passes. The passing grade of 4.0 for our dataset was set at 26 out of the 65 available points, after slight adjustments were made to align with gaps in the distribution of points. The highest grade of 6.0 is calibrated based on point distributions from previous exams. If a student's average grades would fall below a certain threshold requiring a semester repeat, the exam undergoes regrading by a third TA.

Students are aware that they do not need to solve all problems to pass the exam. Due to the stringent time constraints of the exam, many students are unable to address all the items within the given time. Thus, students often choose which problems to prioritize, leaving some items unsolved or incomplete. This results in a sparse dataset, where many student-item combinations are missing, a normal feature in the German university tradition. Despite this, the grading process is fine-tuned to ensure fairness, and students are encouraged to demonstrate understanding through strategic problem selection.

\subsection{Sample}
Studies involving human subjects are reviewed by ETH Zurich's Ethics Committee; this study was approved as Study~2023-N-286. Out of 434 students taking the exam, 252 agreed to participate in the study by completing an informed consent form. The consent form was distributed to all students before the exam and collected at the end, alongside their exam submissions. It remains unclear whether the opt-in informed consent process introduced any bias. Potential biases could have arisen if students disregarded the form due to the high-stakes nature of the exam, chose not to submit it after perceiving poor performance, or simply ran out of time to complete it. Despite these uncertainties, the sample exhibited a wide range of performance. However, due to adherence to the research protocol, we were unable to compare the demographics or performance of participants with non-participants, limiting insights into potential differences between the groups. The 252 students turned in 3041 pages of handwritten exam solutions, an average of 12~pages per student.

\subsection{Cloud Infrastructure}
Access to OpenAI models was provided through Azure AI Services~\cite{azure}, where ETH Zurich has a contract that ensures confidentiality, with a consumption-based per-token payment structure. We used GPT-4o~\cite{gpt4o} for handwriting recognition on US-American data centers (as at the time of our study, the model was not yet available in Europe; students were made aware that data might be processed in the United States in the informed consent form) and  GPT-4-Turbo for reasoning and grading judgement on Swedish data centers. Due to the sensitivity of the data, in addition to the contractual safeguards, we used pseudonyms instead of clear-text names or student identification numbers. Including handwriting recognition and grading, the cost per exam for token usage was approximately \$7.

\subsection{Ground Truth}
The TA-assigned scores per rubric item were taken as ground truth. With very few exceptions, the TAs gave either full or no credit for a particular rubric item. However, these data were not referred to until after all grading and statistical analyses were completed.

\section{Methodology}
\subsection{Workflow}
\begin{figure*}
\begin{center}
\includegraphics[width=\textwidth]{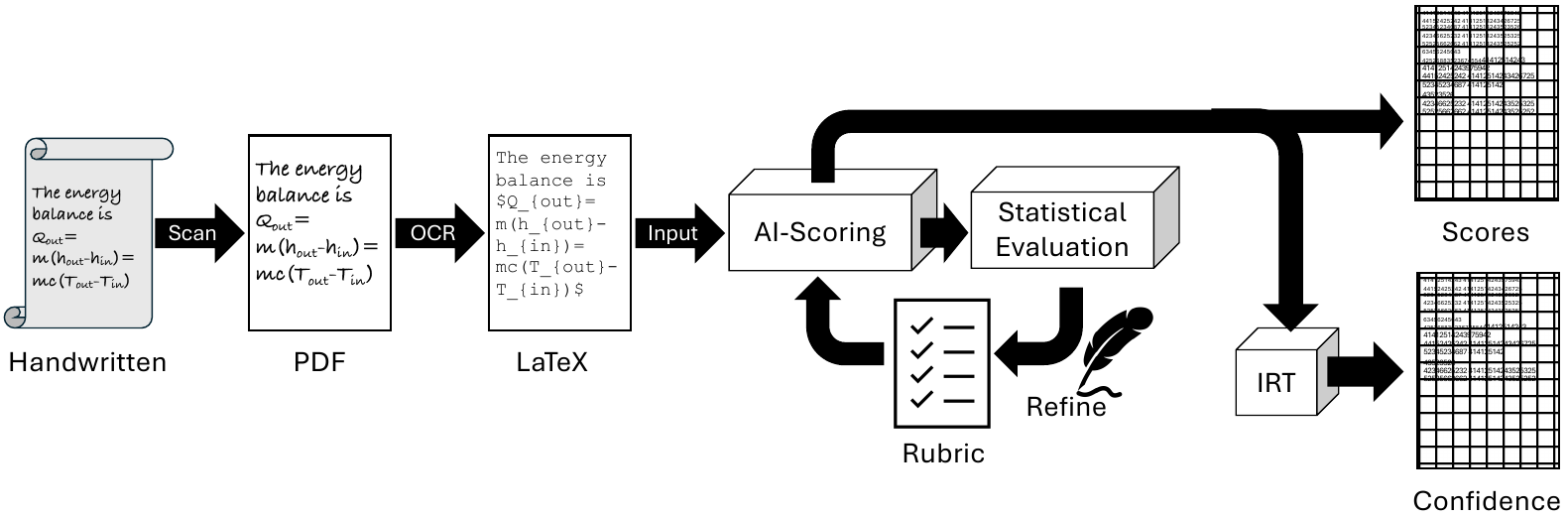}
\end{center}
\caption{Workflow of exam grading, starting with scanning the exam sheets, making them machine-readable using GPT-4-o, evaluating them using GPT-4-Turbo and {\tt R} packages (starting with an initial rubric and going through refinement iterations), and establishing confidence measures using Python code.}
\label{fig:workflow}
\end{figure*}

Figure~\ref{fig:workflow} shows the workflow employed in this study. The handwritten exams were first scanned to PDF using a copy machine with automatic paper feed. The PDFs were then converted to LaTeX using GPT-4o as an OCR tool.

The LaTeX files of the student solution were afterwards graded one problem part at a time, following the results of an earlier exploration~\cite{kortemeyer2024grading}. The part number to be graded, the problem text, the rubric items, and the student solution were used to generate the grading prompts. After initially grading all problem parts using ten independent runs of GPT-4-Turbo, a statistical evaluation was carried out using standard deviations, IRT, and correlations. Calculations are performed using the {\tt qgraph}~\cite{qgraph} and Latent Trait Model ({\tt ltm})~\cite{ltm} packages within the {\tt R} statistical software system~\cite{rpackage}.

Based on the result from the statistical evaluation the prompts were iteratively adjusted using descriptive and Bayesian statistics. 
If a grading category was rewritten, the problem part which contained the rule was regraded using AI. After the grading with the final set of rubric items, problem parts for which no student work could be found were filtered, and the resulting sparse dataset was analyzed using a customized Python program for IRT-estimates~\cite{kortemeyer2019quick} as well as {\tt Scikit}~\cite{pedregosa2011scikit} and {\tt Pandas}~\cite{mckinney2011pandas}.

\subsection{Optical Character Recognition}
For the Optical Character Recognition (OCR), GPT-4o was used the following role,
\begin{quote}\scriptsize
Your role is to interpret images from handwritten exams, converting them into LaTeX format where possible or providing detailed descriptions in German if they represent drawings or figures. Use your capabilities to ensure accuracy and clarity in the translations or descriptions.
\end{quote}
and the following prompt,
\begin{quote}\scriptsize
Please convert the following image of a page out of a handwritten student solution to a physics exam to LaTeX.

Output a LaTeX document that contains all the text and formulas, as well as very detailed verbal descriptions of all graphs, figures, and diagrams.

The text and formulas need to be faithful representations of what the student wrote.

The verbal descriptions of graphical content need to allow a reader of the LaTeX document to reconstruct what the graphical content looks like,
so it needs to verbally describe in detail the graphs and axes, etc. Do not use TikZ, just words. 

Do not comment on anything, output only the utf-8 encoded LaTeX that carefully and closely represents the student solution. 
\end{quote}
The exam solutions were converted as individual page images, which were supplied as Base64-encoded data, and then recompiled into one LaTeX file.

\subsection{Rubric Items}\label{sec:rubric}
\begin{figure*}
\begin{center}
\includegraphics[width=\textwidth]{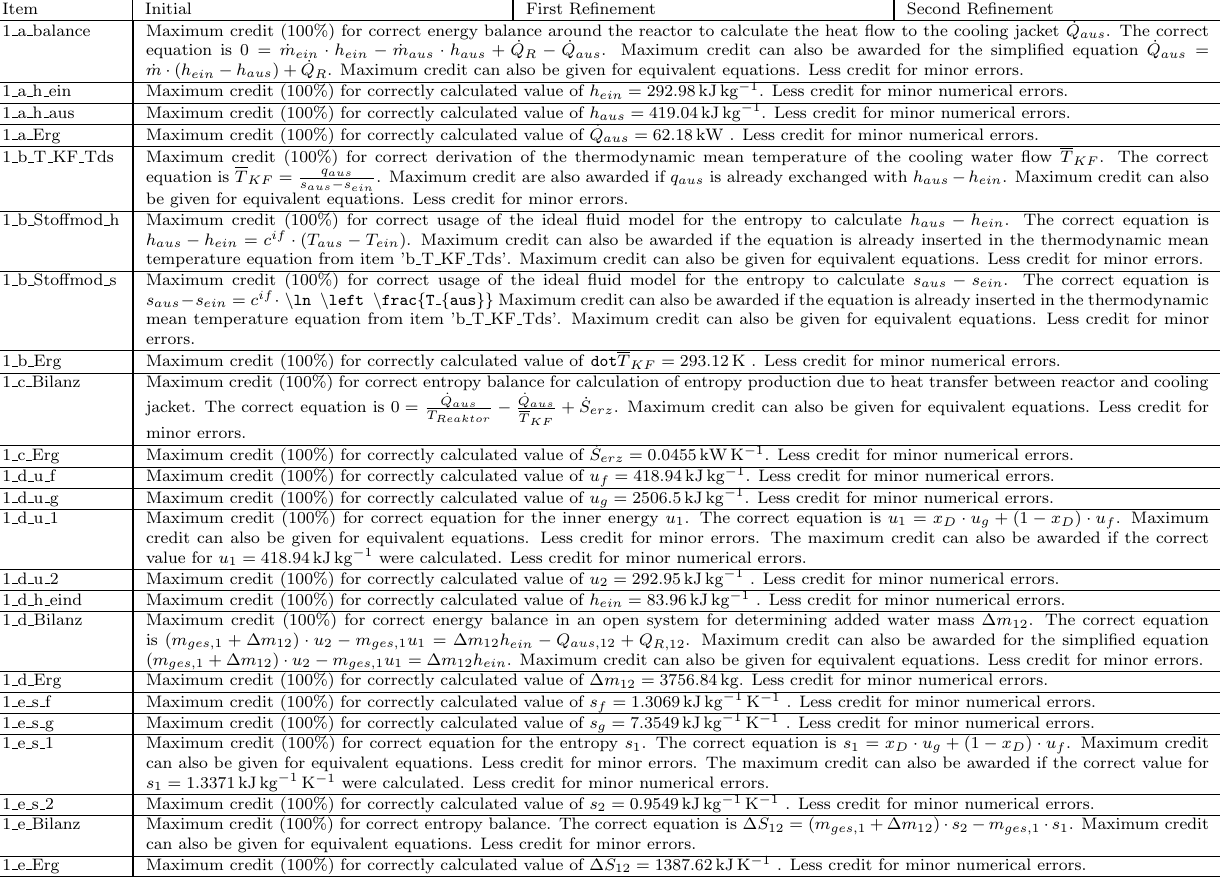}
\end{center}
\caption{Rubric items for problem~1. These items were not refined between iterations, and copy/paste errors in rubric items 1\_b\_Stoffmod\_s an 1\_b\_Erg remained unnoticed.}
\label{fig:rp1}
\end{figure*}

\begin{figure*}
\begin{center}
\includegraphics[width=\textwidth]{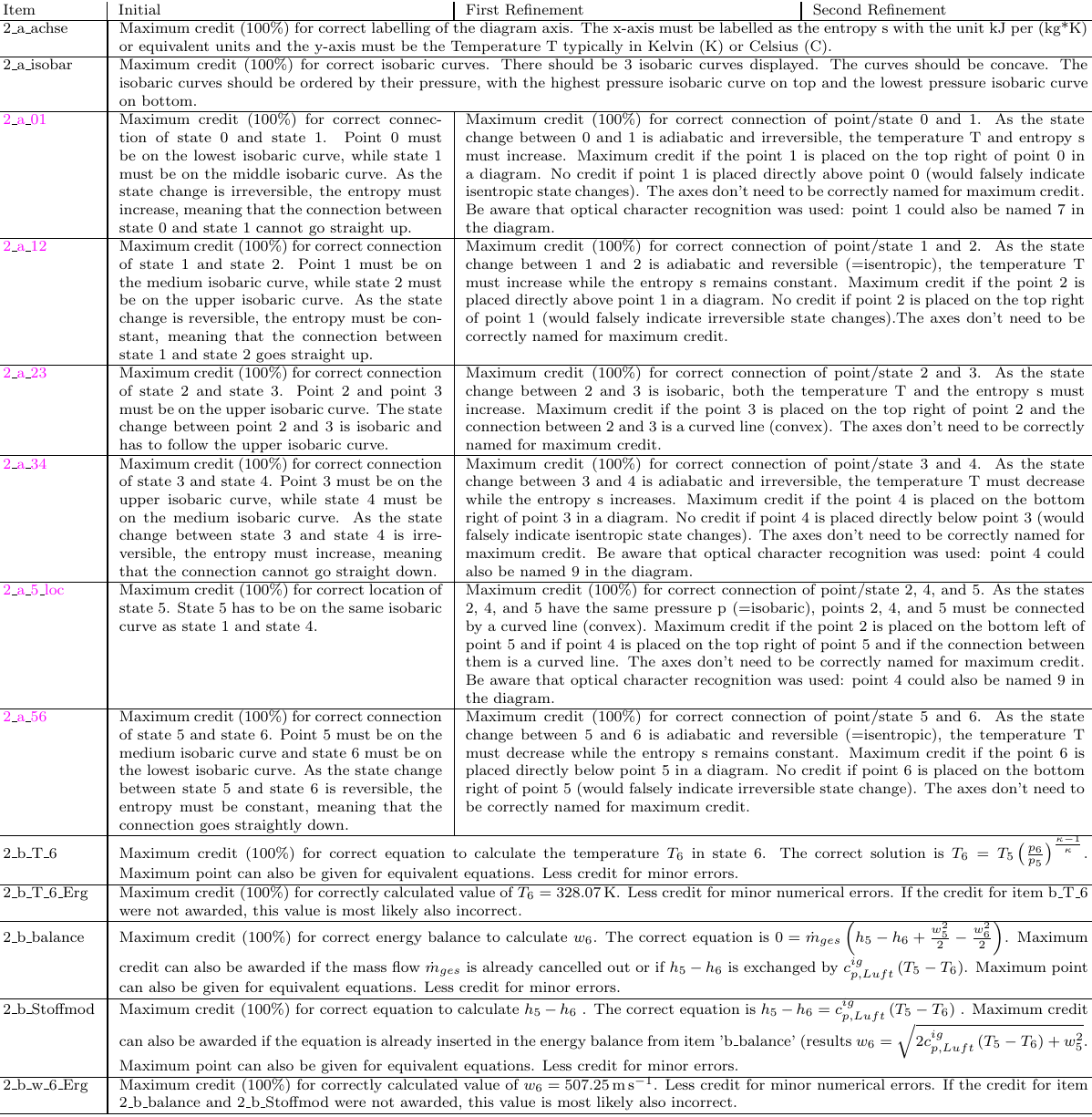}
\end{center}
\caption{Rubric items for problem~2, parts~a and~b. Six of the rubric items were refined in the first iteration.}
\label{fig:rp2ab}
\end{figure*}

\begin{figure*}
\begin{center}
\includegraphics[width=\textwidth]{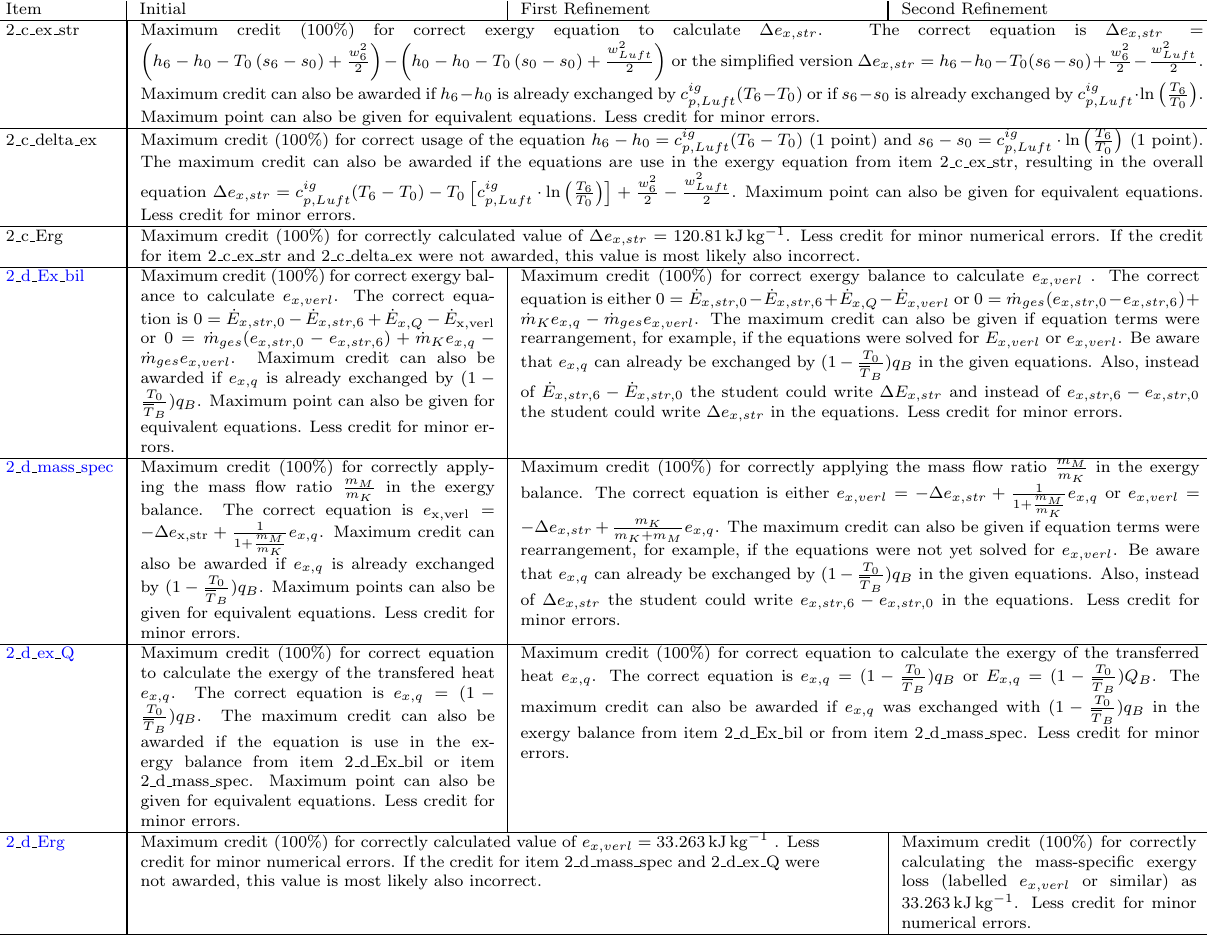}
\end{center}
\caption{Rubric items for problem~2, parts~c and~d. Three rubric items were refined in the first iteration, and 2\_d\_Erg in the second iteration.}
\label{fig:rp2cd}
\end{figure*}

\begin{figure*}
\begin{center}
\includegraphics[width=\textwidth]{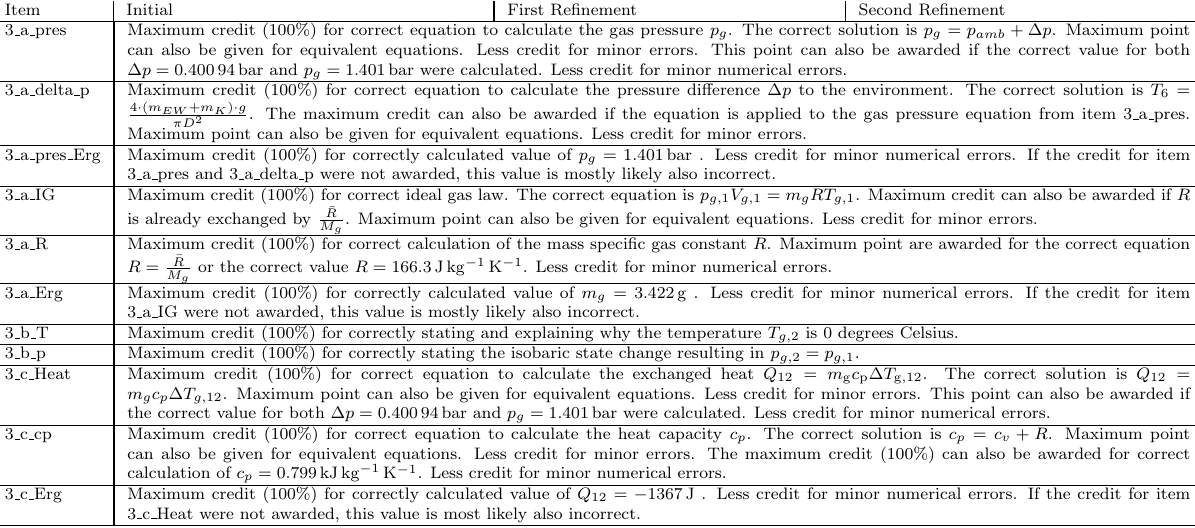}
\end{center}
\caption{Rubric items for problem~3, parts~a,~b, and~c. These rubric items remained unchanged. The model would need to base the correctness of 3\_b\_T on the problem statement and its own training, as the sample solution was not provided.}
\label{fig:rp3abc}
\end{figure*}

\begin{figure*}
\begin{center}
\includegraphics[width=\textwidth]{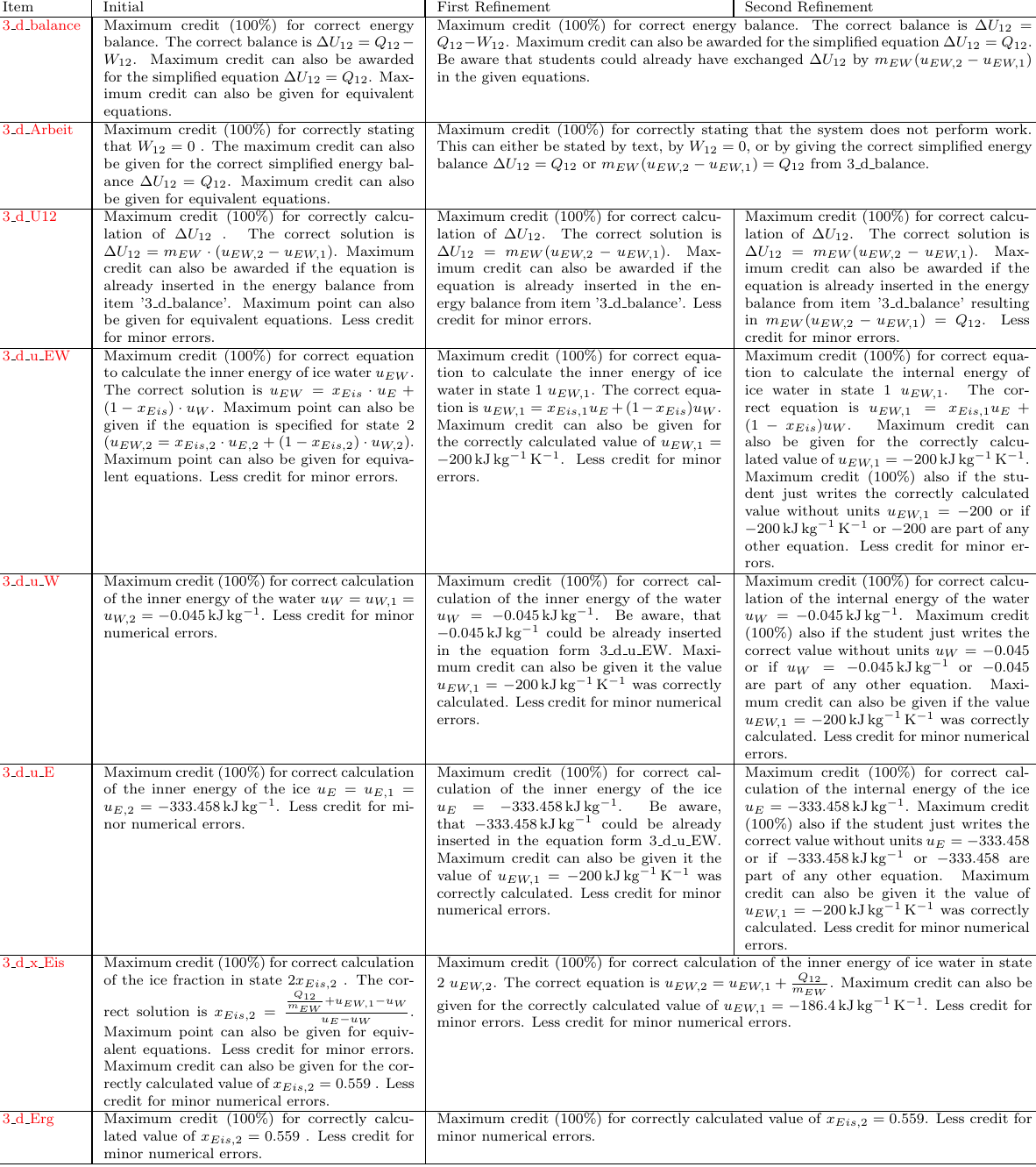}
\end{center}
\caption{Rubric items for problem~3, part~d. All rubric items were refined in the first iteration, and four were refined again in the second iteration.}
\label{fig:rp3d}
\end{figure*}

\begin{figure*}
\begin{center}
\includegraphics[width=\textwidth]{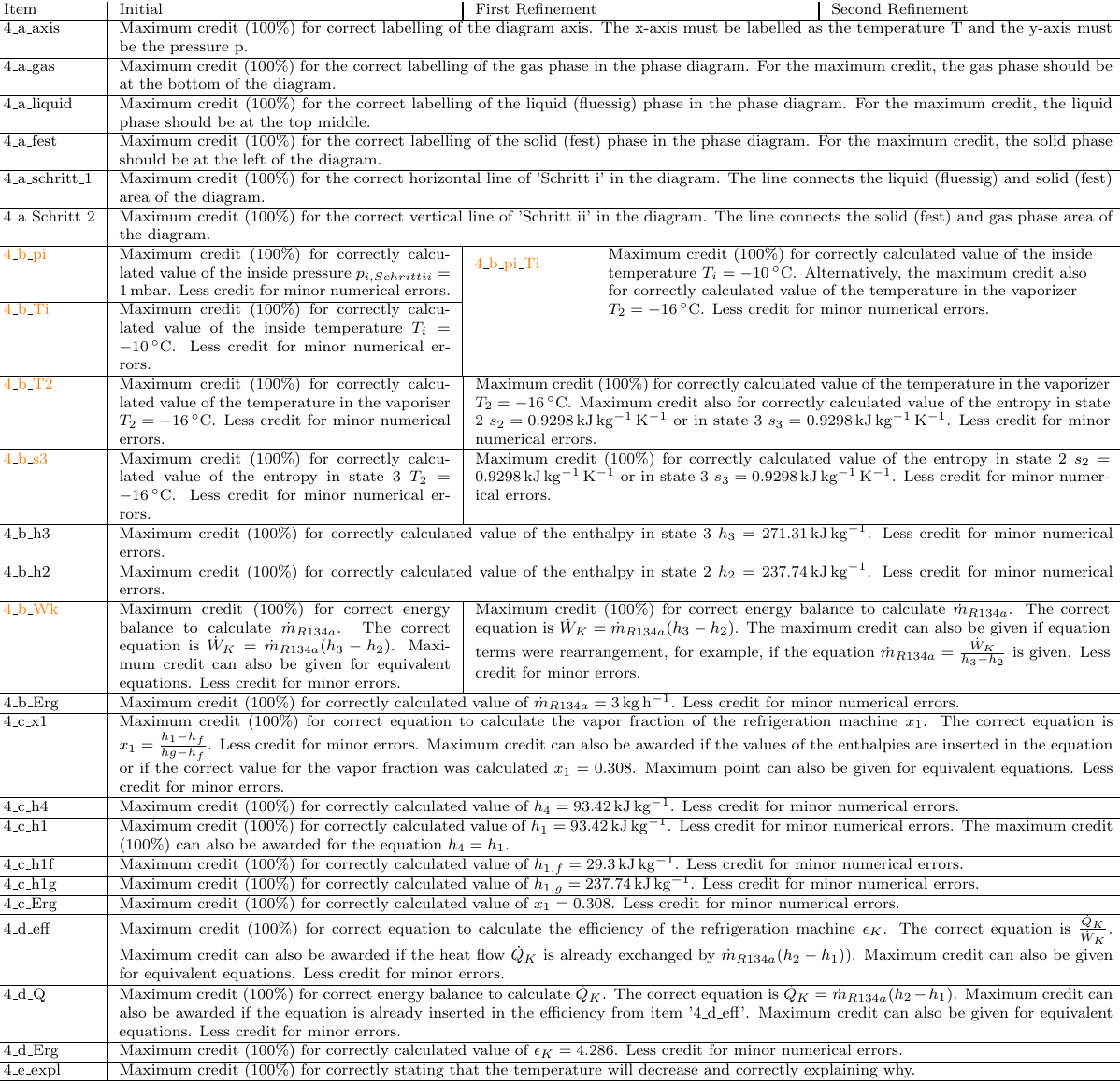}
\end{center}
\caption{Rubric items for problem~4. Five of the rubric items were refined in the first iteration, and in particular, rubric items 4\_b\_pi and 4\_b\_Ti were combined. The model would need to base the correctness of 4\_e\_expl on the problem statement and its own training, as the sample solution was not provided.}
\label{fig:rp4}
\end{figure*}

The exams were graded by teaching assistants based on a sample solutions that had point values attached to the steps of the calculations.These were transcribed into English and separate rubric items, which were labelled using unique identifiers such as ``2\_a\_isobar'' for drawing the correct isobaric curve in the $T-s$ diagram for the engine in part~a of problem~2~\footnote{The identifiers were intended for internal use and are a mixture of English and German. For example, ``Erg'' is short for ``Ergebnis,'' which is German for ``result.''}. The rubric items, shown~\Cref{fig:rp1,fig:rp2ab,fig:rp2cd,fig:rp3abc,fig:rp3d,fig:rp4}, in turn were submitted as the prompt for each problem part for each student to the AI system. While the students' exam solutions were generally submitted in German, in earlier explorations we had found anecdotal evidence that at least earlier versions of the LLM appeared to reason better in English than in German. Also based on earlier experiments, the system was prompted to grant partial credit between 0\% and 100\%, even though the TAs generally tended to make binary decisions, that is, full credit or none. Prompting the LLM differently will allow us to set a correctness threshold $C$ later on.

\subsection{Grading}
For the grading step, GPT-4-Turbo was used with the following role,
\begin{quote}\scriptsize
You are a teaching assistant for a university course on thermodynamics, tasked with assisting in grading exams.

The course is taught in German, and the exams are also written in German. You will receive OCR-scanned, originally handwritten student solutions, which might contain scanning errors, especially in symbols and equations.

Your task is to interpret these solutions accurately, ensuring that percentage credit is awarded only for work that meets the rubric's criteria.

You must carefully grade the student solutions according to the rubric, paying close attention to the presence and quality of the work.
\end{quote}
and the following prompt template,
\begin{quote}\scriptsize
You are tasked with grading part [partnumber] of the student solutions for this problem:

[problem]

==============

Below is the grading rubric table in tab-separated format. The first column contains the problem part, the second column provides the rubric item identifier, and the third column contains the grading criterion.

[rubric]

==============

The problem has multiple parts, but you only grade part [partnumber]. There are several rubric items that are graded according to criteria. Note that likely students will not have worked on all parts and items, and they may not have done work in the same order as the rubric lists them.

This is what you need to do:

* Go through every row in the grading rubric table, do not skip any. The row is identified by the rubric item identifier listed in the second column of the rubric table, and that is how you remember it. 

* For each row, check if any work pertaining to the criterion is present. If there is no related work, give zero credit (0\%) for that rubric item.

* For each row, if you find work that pertains to the criterion, check how well it corresponds to the criterion and assign partial percentage credit between 0\% and 100\%.

Note that there can be OCR errors, leading to some wrong digits or symbols; this is not the student's fault, and you need to determine if errors are likely due to OCR (no credit deduction) or due to the student (some credit deduction).

Give 100\% if the work fulfills the criterion very well. Give less partial percentage credit the less it agrees. Do not give bonus points. 

* For each row, provide a comment explaining the rationale behind your grading decision.

Output your grading in a three-column csv-separated table format with a row for every item in the rubric. Use newline as the row separator:

The first column in your table must be the rubric item identifier, 
the second column the percentage credit that you awarded (without the percentage sign), and 
the third column your comments explaining the grading decisions.

Output nothing else but this completed grading table, and do not enclose it in any special characters, so it is a syntactically correct csv-separated file.

Here's the student solution that you need to grade:

[solution]

\end{quote}
The terms in square brackets are placeholders for the part number to be graded ({\tt[partnumber]}), the problem text ({\tt[problem]}) and the student solution({\tt[solution]}) for all parts, as well as the rubric items for only the part that should be graded ({\tt[rubric]}). In contrast to our earlier study, the sample solution is not given in the prompt anymore~\cite{kortemeyer2024grading}. It is important to point out that the problem text was part of the system prompt, thus grading would not be based solely on the rubric items.

\subsection{Rubric Refinement}
For the refinement of the rubric, we are looking at three different measures:
\begin{itemize}
\item As each item is graded ten times, the standard deviation of the assigned partial-credit percentages could be considered as a measure of the clarity of the rubric item. The measure is similar to inter-rater reliability, however, in this context is used as a proxy for the clarity of the rubric item; the higher the standard deviation, the more room the rule might leave for interpretation.
\item The item parameters of the rubric items. Assuming content and construct validity of the exam, we are looking for outliers that might be the result of badly written rubric items, for example, items that appear to exhibit extremely high difficulty or negative discrimination might in fact have faulty rubric criteria. As standard 2PL-IRT assumes dichotomous grading decisions, we assumed that 50\% is the threshold above which the
student solution would be considered correct.
\item The correlations between the partial credit that students received on different items might indicate ill-defined rubric items under the assumption that students tend to do generally well on a problem part or not.
\end{itemize}
The formulations of rubric items that based on these three measures were subjectively identified as outliers were refined, and the problem parts that they belonged to were regraded by the AI.

\subsection{Uncertainty Determination}
For the problem parts where the AI identified corresponding work from a particular student, IRT estimates were used to predict the student's performance on the associated rubric items. Specifically, Equation~\ref{eq:2PL} was interpreted to represent the expected partial credit of student $j$ on rubric item $i$. This prediction was based on the student's latent ability parameter $\theta_j$, as well as the item's estimated difficulty $b_i$ and discrimination $a_i$. It assumes that the student's performance is consistent across the problem parts they completed. 

The absolute difference between the predicted partial credit and the AI-assigned score $s_{ij}$ was then taken as a measure of uncertainty $u_{ij}$, that is
\begin{equation}\label{eq:uij}
u_{ij}=\left|p_i(\theta_j)-s_{ij}\right|
\end{equation}
In other words, the greater the discrepancy between the AI-assigned score and the predicted score, the less confidence there is in the grading. Note that up to this point, no reference to ground truth is employed; the matrix $u_{ij}$ is based completely on AI-grading. The mechanism appears similar to a posterior predictive model-checking~\cite{sinharay2006bayesian}, however, here we are not checking the quality of the model but the plausibility of the scores.

An uncertainty threshold $U$ in connection with the boolean parameter $P$ that would only accept problems graded as correct by the AI were then imposed to determine which student-item combinations would be accepted and which should be graded by humans. Mathematically, the boolean acceptance $A_{ij}$ would be given by
\begin{eqnarray}\nonumber
A_{ij}&=&\left\{\begin{array}{ll}
u_{ij}<U&\mbox{if $P$=false}\\
u_{ij}<U \land s_{ij}>C&\mbox{if $P$=true}
\end{array}\right.\\
&=&\left( u_{ij} < U \right) \land \left( \neg P \lor s_{ij} > C \right)
\end{eqnarray}
Thus, the AI-judgement $s_{ij}$ is evaluated and accepted or rejected based on student $j$'s performance on all rubric items and on the performance of all other students on rubric item $i$. 
The fraction of student-items accepted was calculated as a measure of efficiency:
\begin{equation}
\text{Acceptance Rate} = \frac{\sum_{ij} \mathbf{1}_{\{A_{ij}\}}}{\sum_{ij} 1}
\end{equation}

\subsection{Evaluation}
After the final round of rubric refinement and grading, the ground truth was taken into consideration. For the rubric refinement, the agreement between AI and TA grades on total exam points was evaluated using linear regressions, the point values for the AI grading were determined by multiplying the assigned partial credit with the point weighing of the rubric item.

For the evaluation of the grading, for each student, the problem parts for which the AI gave zero credit for all rubric items were taken out of consideration; this accounts for the sparseness of the dataset given that students generally only submitted work for selected problem parts. Also, all student-items for which the AI grading based on Bayesian statistics had an uncertainty larger than a given uncertainty threshold were discarded. For different values of correctness threshold and uncertainty threshold, the evaluation was carried out in two ways: using standard machine-learning measures and linear regression parameters. For the machine-learning measures, one can define the following:
\begin{itemize}
\item True Positive (TP): the student-item graded as correct by the AI is also graded as correct by the TAs.
\item True Negative (TN): the student-item graded as incorrect by the AI is also graded as incorrect by the TAs.
\item False Positive (FP): the student-item graded as correct by the AI is graded as incorrect by the TAs.
\item False Negative (FN): the student-item graded as incorrect by the AI is graded as correct by the TAs.
\end{itemize}
The standard measures accuracy, precision, recall, and F1-score are then defined as follows:
\begin{eqnarray}
\text{Accuracy} & = & \frac{\mbox{TP} + \mbox{TN}}{\mbox{TP} + \mbox{TN} + \mbox{FP} + \mbox{FN}} \\
\text{Precision} & = & \frac{\mbox{TP}}{\mbox{TP} + \mbox{FP}} \\
\text{Recall} & = & \frac{\mbox{TP}}{\mbox{TP} + \mbox{FN}} \\
\text{F1-score}& = & 2\cdot \frac{\text{Precision} \cdot \text{Recall}}{\text{Precision} + \text{Recall}}
\end{eqnarray}
Accuracy reflects how often the AI's grading matches the grading done by teaching assistants (TAs); it considers both correct and incorrect classifications, showing the overall effectiveness of the AI across all graded items.
Precision measures how reliable the AI is when it assigns a correct grade; it tells us the proportion of AI-graded correct answers that actually match the TAs' correct grades, minimizing false positives.
Recall (or sensitivity) indicates how well the AI identifies correct answers, showing the proportion of TA-correctly-graded items that the AI also grades as correct, minimizing false negatives.
The F1-score is the harmonic mean of precision and recall, balancing the two measures to provide a single metric that reflects both the AI's precision in avoiding false positives and its ability to catch true positives; this score is especially useful when there is an imbalance between false positives and false negatives.

For the linear regression of AI versus TA grades, perfect agreement would signaled by a slope of one, an offset of zero, and a coefficient of determination $R^2$ of one. Different offsets could signal that the AI is erroneously giving away or withholding credit on certain items, and a lower $R^2$ would indicate more randomness in the AI grading.

\section{Results}
\subsection{Refinement of the Rubric}\label{sec:refine}
\begin{figure*}
\begin{center}
\includegraphics[width=0.329\textwidth]{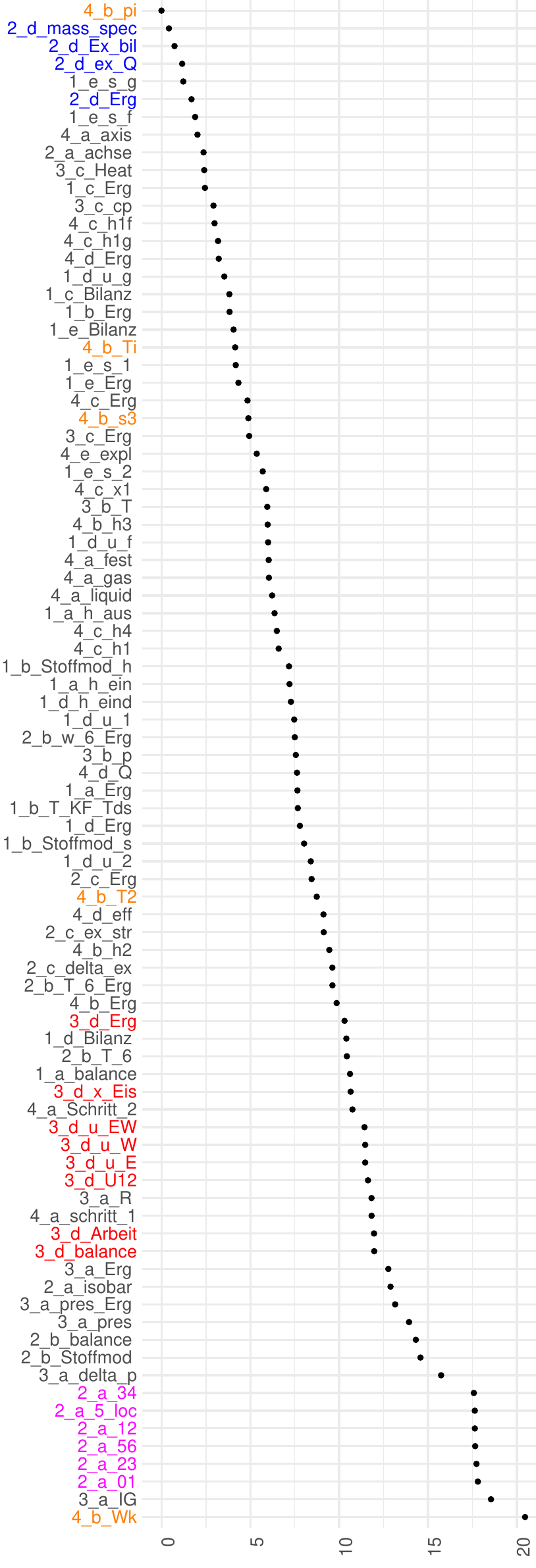}
\includegraphics[width=0.329\textwidth]{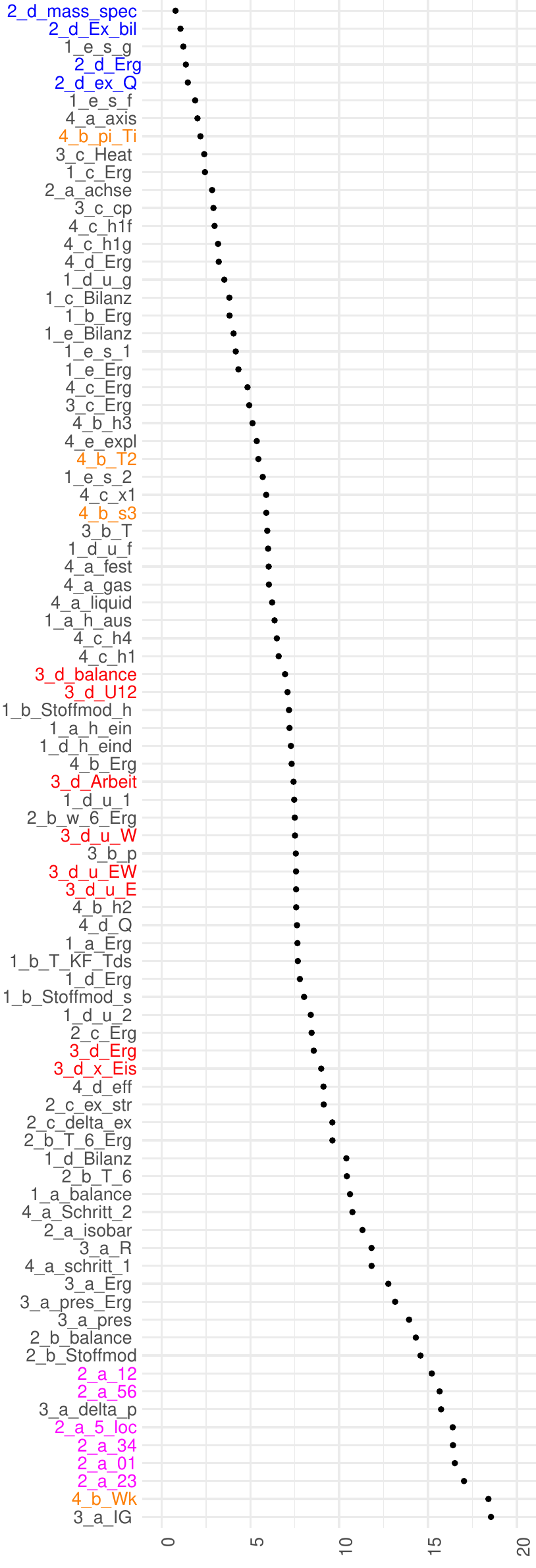}
\includegraphics[width=0.329\textwidth]{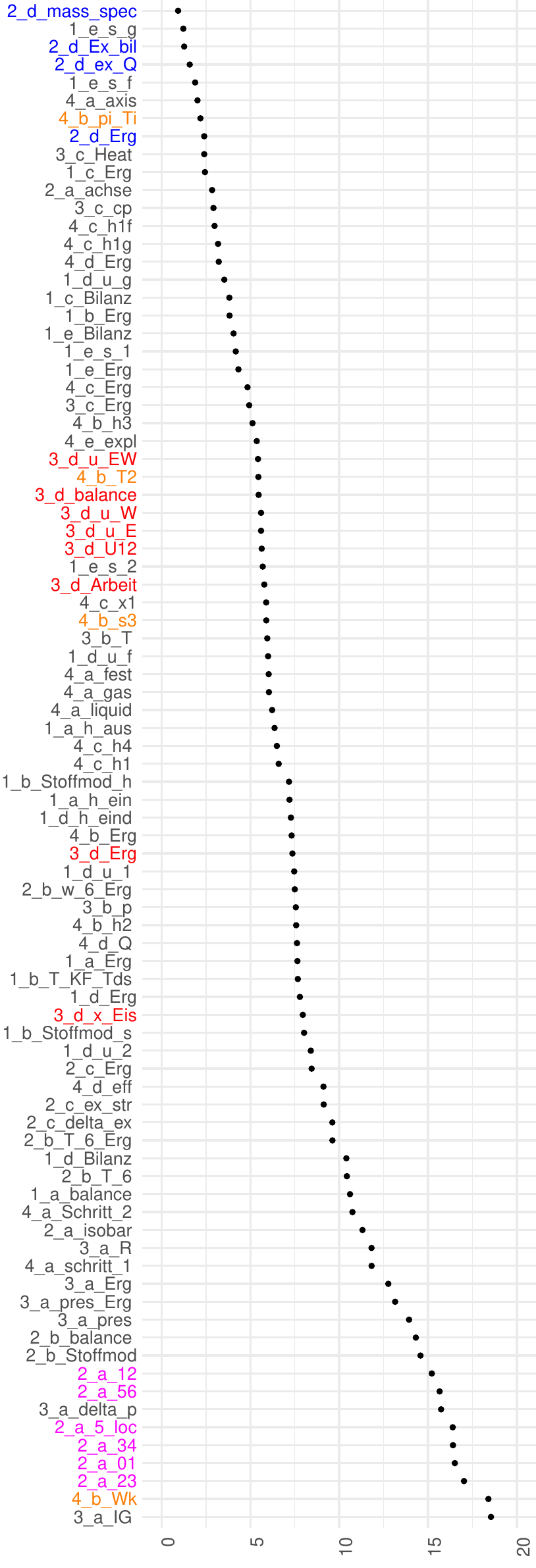}

{\sf Standard Deviations of AI-Item Scores (in percent)}
\end{center}
\caption{Sorted plot of the standard deviations of the average AI-scores for each rubric item, resulting from left to right from the initial grading, as well as the first and second round of refining the rubric. Modified rubric items have been color-coded according to problem part: part 2a in magenta (see Fig.~\ref{fig:rp2ab}), 2d in blue (Fig.~\ref{fig:rp2cd}), 3d in red (Fig.~\ref{fig:rp3d}), and 4b in orange (Fig.~\ref{fig:rp4}).}
\label{fig:stdev}
\end{figure*}

\begin{figure*}
\begin{center}
\includegraphics[width=0.329\textwidth]{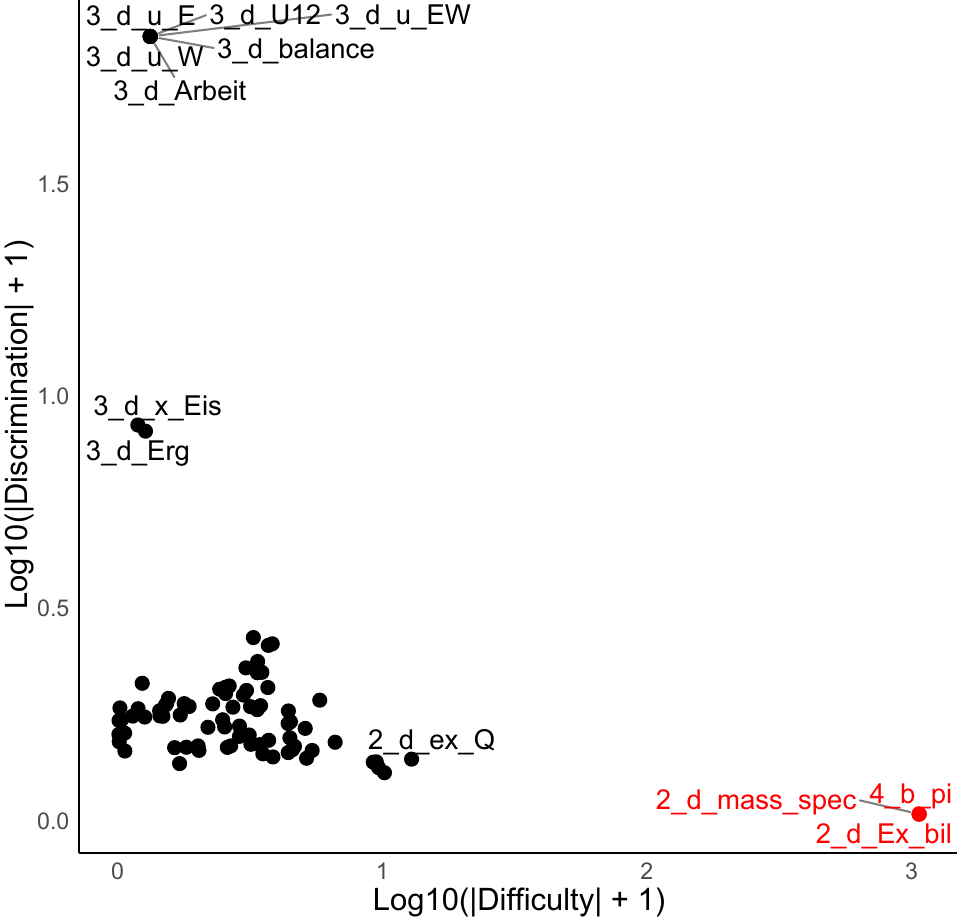}
\includegraphics[width=0.329\textwidth]{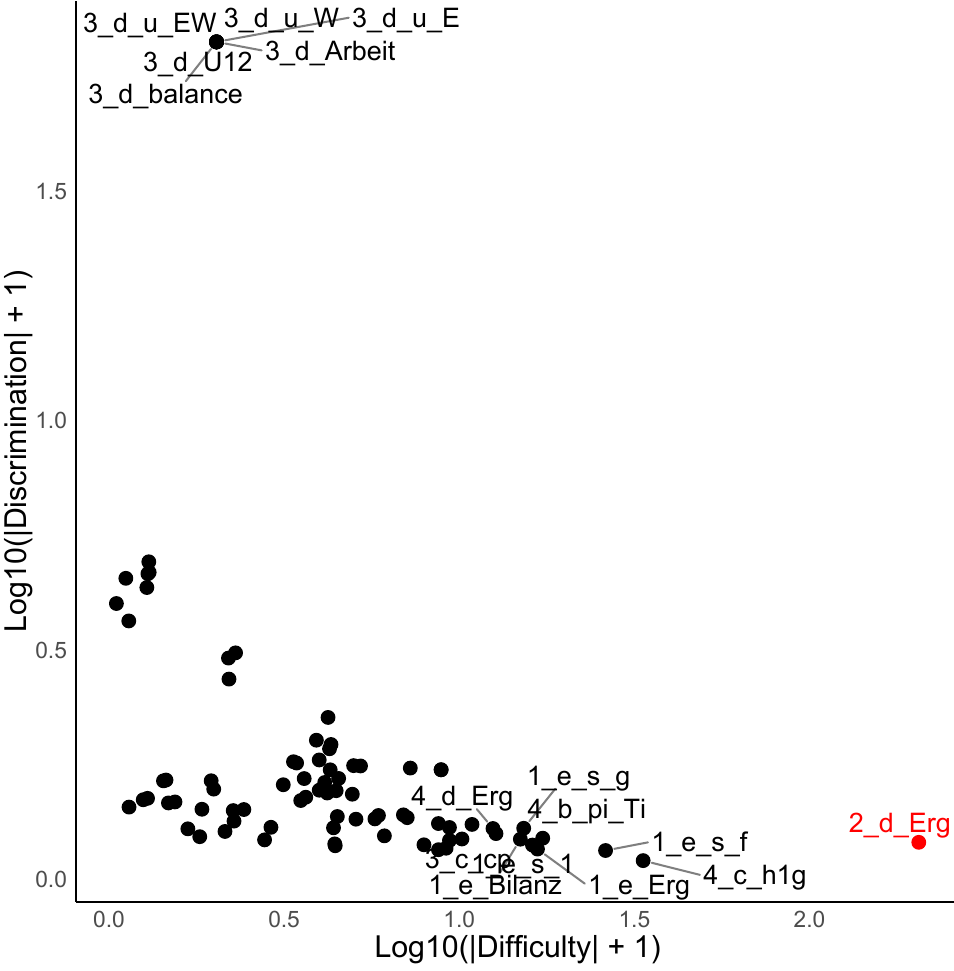}
\includegraphics[width=0.329\textwidth]{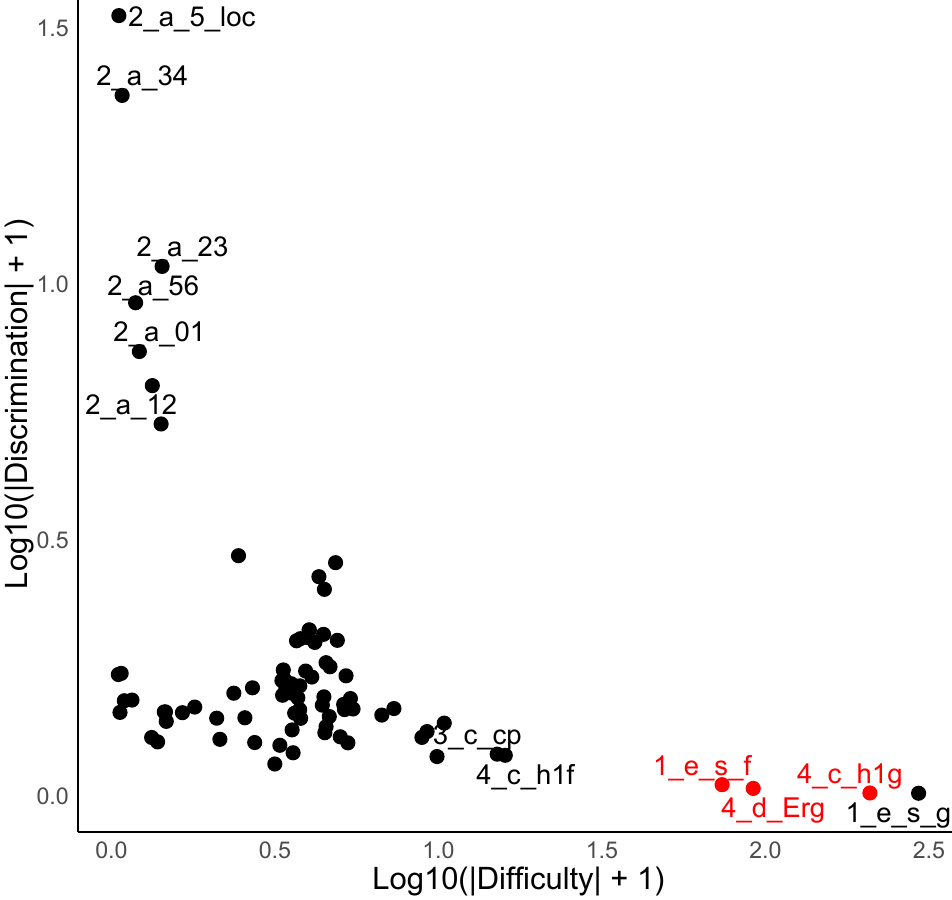}
\end{center}
\caption{Item Response Theory parameters of the rubric items, resulting from the initial grading (see also Fig.~\ref{fig:irtcurves}), as well as the first and second round of refining the rubric. Negative values of the discrimination are indicated by red markers.}
\label{fig:irt}
\end{figure*}

\begin{figure*}
\begin{center}
\includegraphics[width=0.65\textwidth]{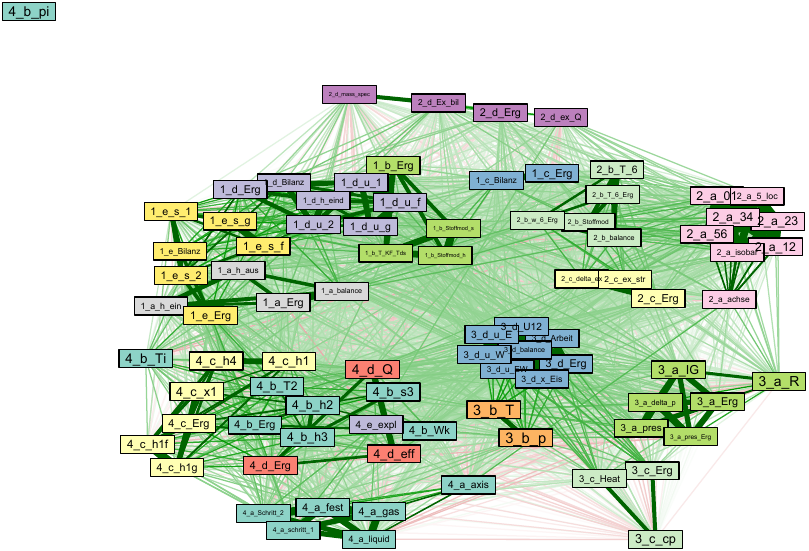}

\includegraphics[width=0.52\textwidth]{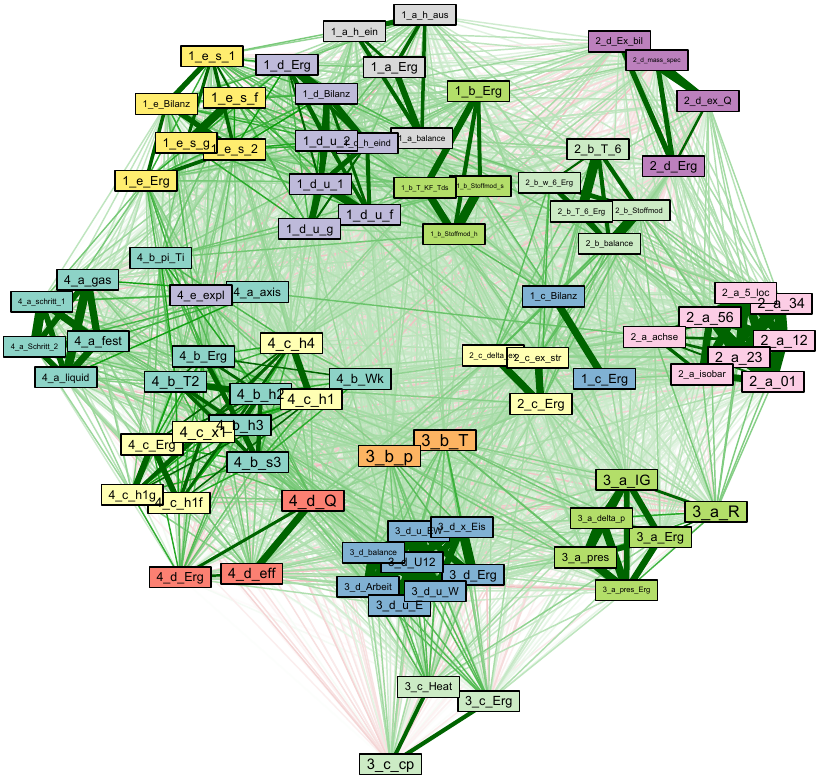}
\includegraphics[width=0.472\textwidth]{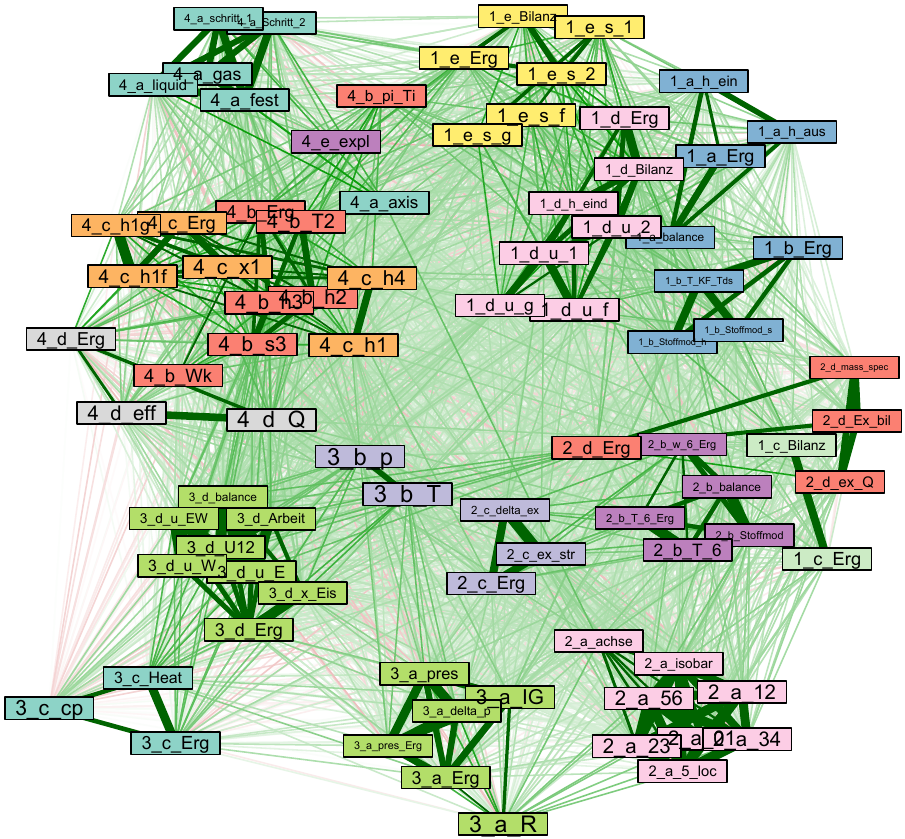}
\end{center}
\caption{Fruchterman-Reingold~\citep{fruchterman1991,qgraph} representations of the correlations between the rubric items, resulting from the initial grading in the top row, as well as the first and second round of refining the rubric on the left and right of the bottom row, respectively. Orientation and handedness of the graphs are random; within each graph, same coloration denotes the same problem part.}
\label{fig:correlations}
\end{figure*}

After the initial grading, several items were identified as outliers:
\begin{itemize}
\item The majority of the rubric items for Problem~2a exhibited large standard deviations, see the leftmost panel of Fig.~\ref{fig:stdev}. This was not surprising, as students had to draw a process diagram, and grading would be based on a textual description of the graphical features of this diagram; the rules were refined as shown in Fig.~\ref{fig:rp2ab}. Earlier experiments with MathPix~\cite{mathpix} had also shown that the numbers~1 and~4, when written in German-style handwriting, tended to be interpreted as~7 and~9, respectively~\cite{kortemeyer2024grading,liu2024ai,kortemeyer2024ethel}.
\item Two of the rubric items of Problem~2d exhibited negative discrimination, see the leftmost panel of Fig.~\ref{fig:irt}. These items and related items within that problem part were rewritten as shown in Fig.~\ref{fig:rp2cd}.
\item More than half the rubric items in Problem~3d exhibited exceptionally high discrimination while having low difficulty, see the leftmost panel of Fig.~\ref{fig:irt}, as well as the extremely steep Item Characteristic Curves in Fig.~\ref{fig:irtcurves}. The items were rewritten as shown in the middle column of Fig.~\ref{fig:rp3d}.
\item Finally, in Problem~4b, item 4\_b\_pi appeared as an outlier in the top panel of Fig.~\ref{fig:correlations}. It turned out that the AI graded none of the student answers as correct. The decision was made to combine this item with the closely related item 4\_b\_Ti, as can be seen in Fig.~\ref{fig:rp4}.
\end{itemize}
Problems~2a,~2d,~3d, and~4b were subsequently regraded by the AI, leading the middle panels in~\Cref{fig:stdev,fig:irt,fig:correlations}.

Unfortunately, while the item parameters of the other rubric items in Problem~2d improved, the formerly unmodified item 2\_d\_Erg exhibited negative discrimination. Also, several items of Problem~3d remained outliers, leading us to refine the corresponding rubric items in the rightmost columns of Figs.~\ref{fig:rp2cd} and~\ref{fig:rp3d}; subsequently, Problems~2d and~3d were regraded by the AI. As the rightmost panel of Fig.~\ref{fig:irt} shows, this fixed some issues but introduced apparent new issues with Problems~1e,~4c, and~4d. It was, however, decided to abort further refinement of the rubric items at this point.

\subsection{Evaluation}
We now introduce the ground truth to evaluate the rubric refinement process and effect of threshold parameters for grading.
\subsubsection{Evaluation of the Rubric Item Refinement}
\begin{figure}
\begin{center}
\includegraphics[width=\columnwidth]{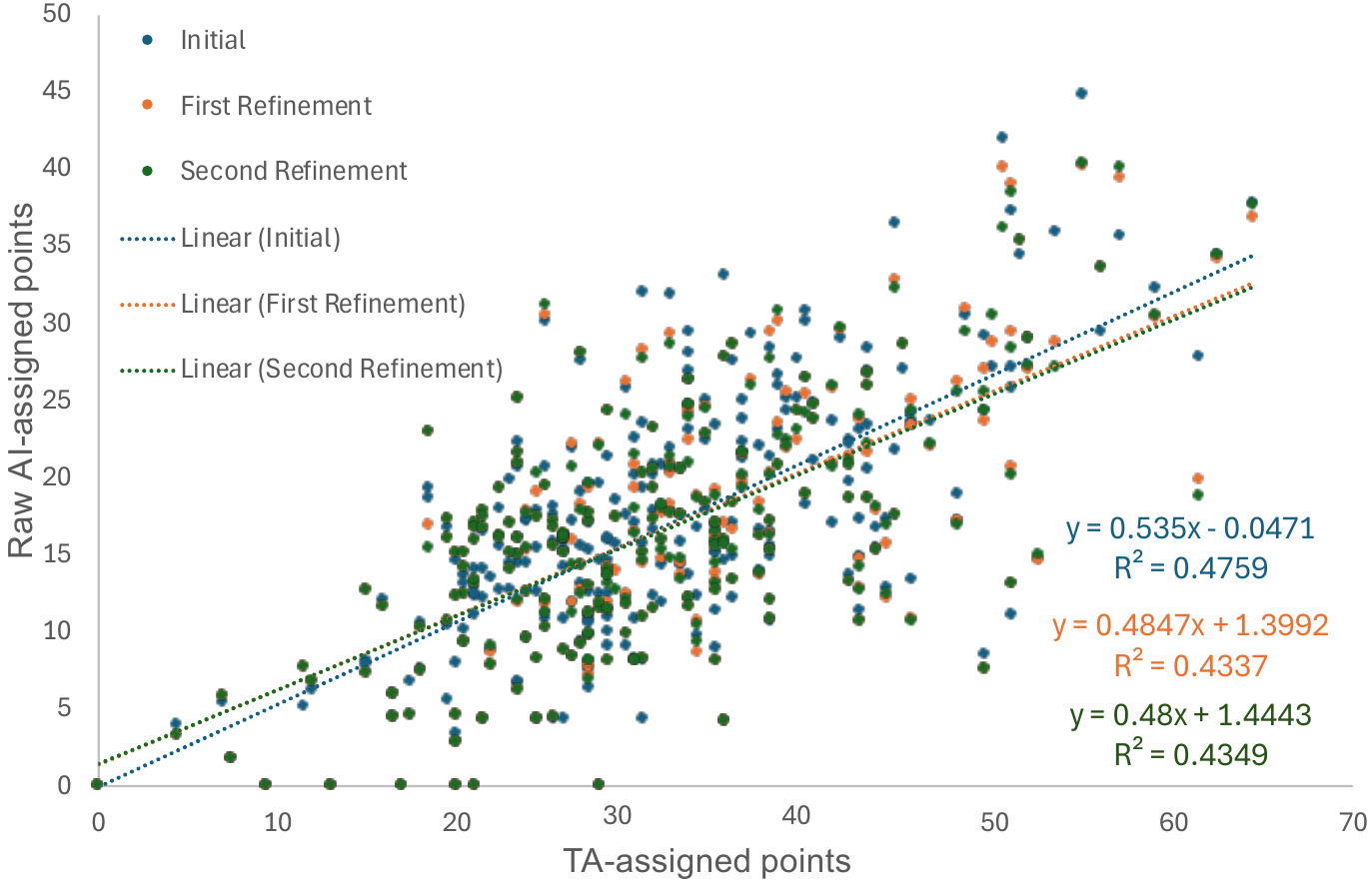}
\end{center}
\caption{Raw AI-assigned versus TA-assigned points for the complete exa, comparing the outcomes for the iterations of rubric refinement discussed in Sect.~\ref{sec:refine}. Each data point represents one student; shown also are the linear regression lines and equations.}
\label{fig:iterations}
\end{figure}

Figure~\ref{fig:iterations} shows the relationship between the TA-assignment points and the raw AI-assigned points for the complete exam for the iterations of the rubric refinement; no thresholds were applied, and all problem parts, including those that a student did not work on, were considered.

The effect of rubric refinement was unexpectedly small, and in fact, led to a slight decrease in model performance: the slope (regression coefficient) decreased even further below 1.0, the offset (intercept) increased, and the coefficient of determination decreased between iterations.  Also, copy/paste errors that occurred by transcribing the TA's sample solution into grading rules, such as they occurred in items 1\_b\_Stoffmod\_s and 1\_b\_Erg (see Fig.~\ref{fig:rp1}), were not detected by the statistical methods employed. Also, items 3\_b\_T and 4\_e\_expl, which simply stated ``correctly explain" without giving an example of a correct explanation remained inconspicuous.

\subsubsection{Evaluation of the Confidence Parameters}
\begin{figure*}
\begin{center}
\includegraphics[width=0.88\textwidth]{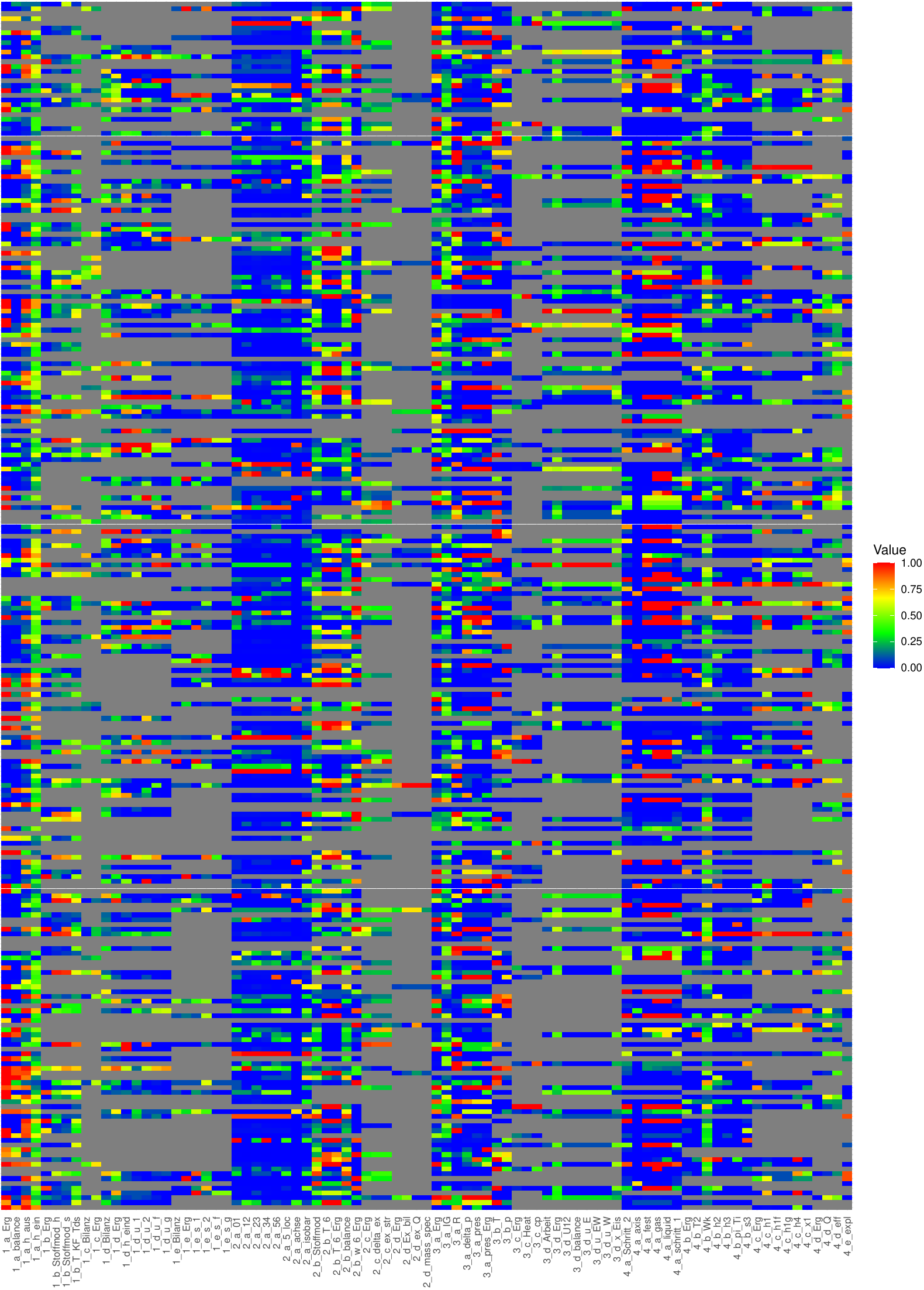}
\end{center}
\caption{The uncertainty matrix $u_{ij}$ (Eq.~\ref{eq:uij}). The matrix shows the rubric items in its columns and the students in its rows. Values between $0.0$ and $1.0$ are indicated by colors shown in the legend, while problem parts for which no work was found are indicated in grey. The problematic items 1\_b\_Stoffmod\_s, 1\_b\_Erg, 3\_b\_T, and 4\_e\_expl leave no clear signature in this matrix, possibly due to the fact that no many students worked on the associated problem parts.}
\label{fig:absdiff}
\end{figure*}

The rather dismal regression model parameters of the raw, unfiltered AI-grading  in Fig.~\ref{fig:iterations} underline the need to scrutinize the AI-results beyond mere adjustments of the grading criteria.
Figure~\ref{fig:absdiff} shows the uncertainty matrix $u_{ij}$. The large grey areas indicate the problem parts that students skipped, blue indicates low uncertainty, and red high uncertainty. For all problems, students tended to skip the later problem parts. The matrix indicates surprisingly high certainty for the graphical Problem~2a, and also considerable certainty for Problems~3a and 4a\&b. Consistently high uncertainty is associated more strongly with particular students than particular rubric items. In this context, it is worth remembering that the uncertainty is the compound of the OCR and rubric-grading uncertainties, and the student-specific uncertainty may well be the result of unclear handwriting.

\begin{figure*}
\begin{center}
\includegraphics[width=0.49\textwidth]{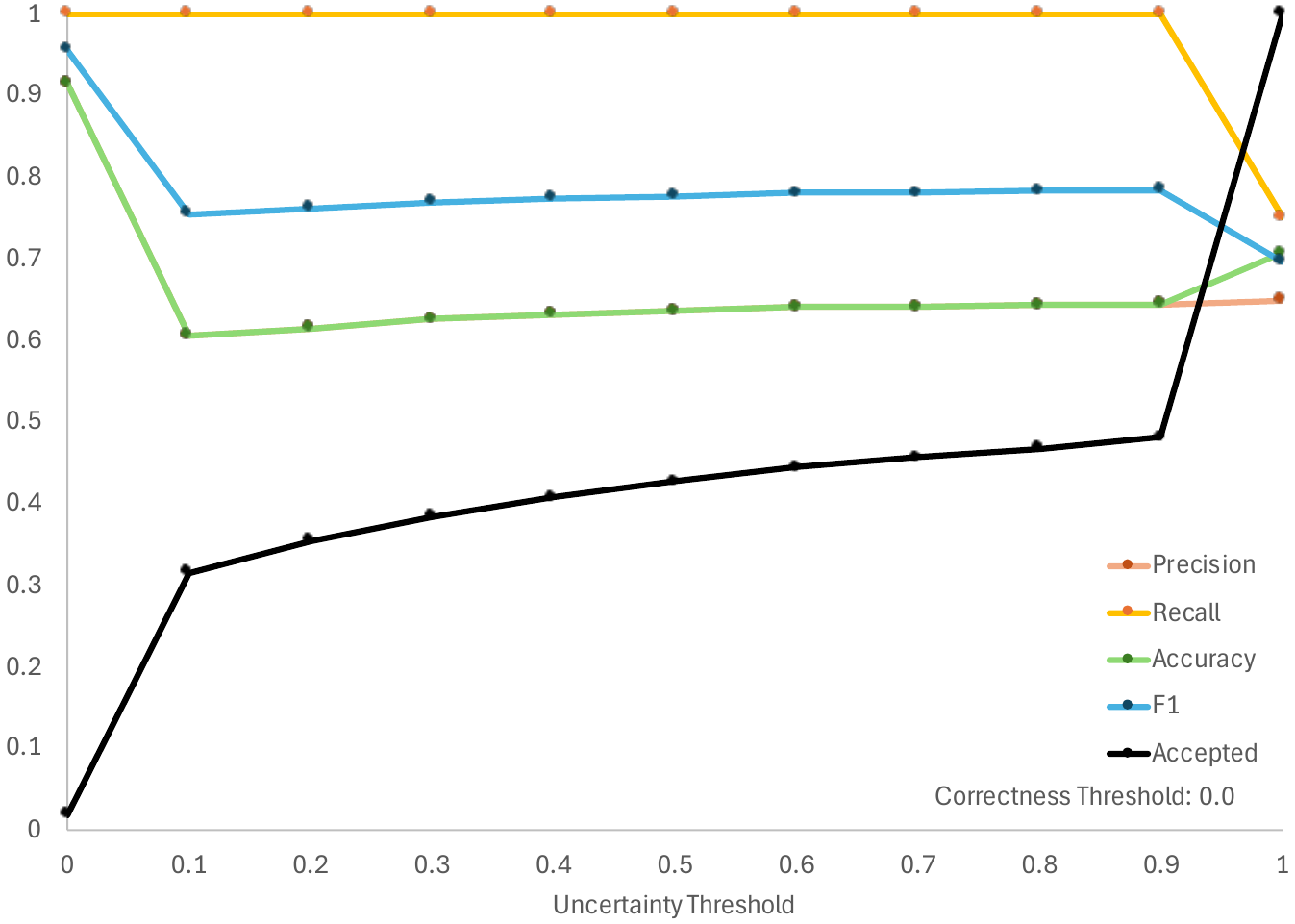}
\includegraphics[width=0.49\textwidth]{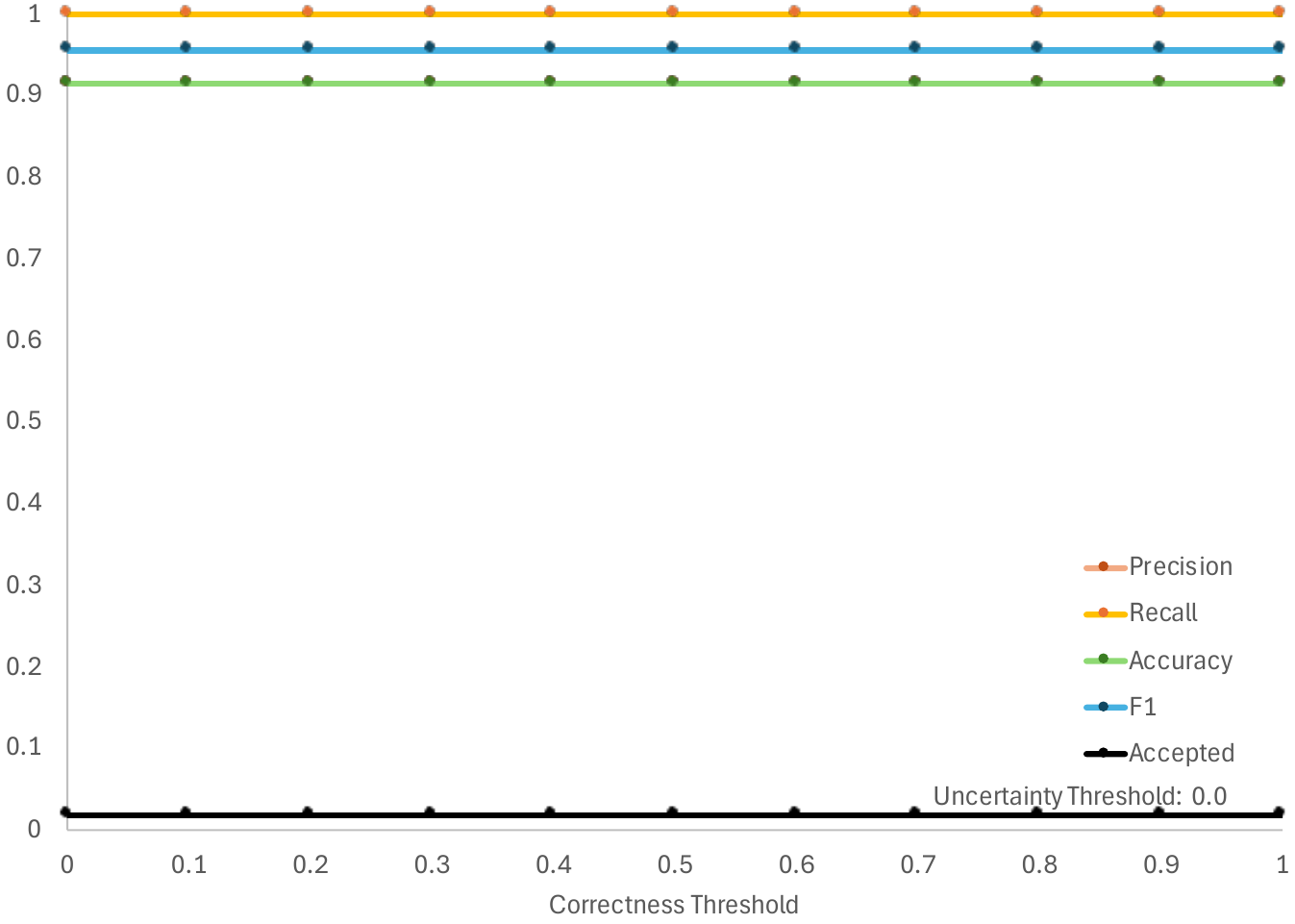}

\includegraphics[width=0.49\textwidth]{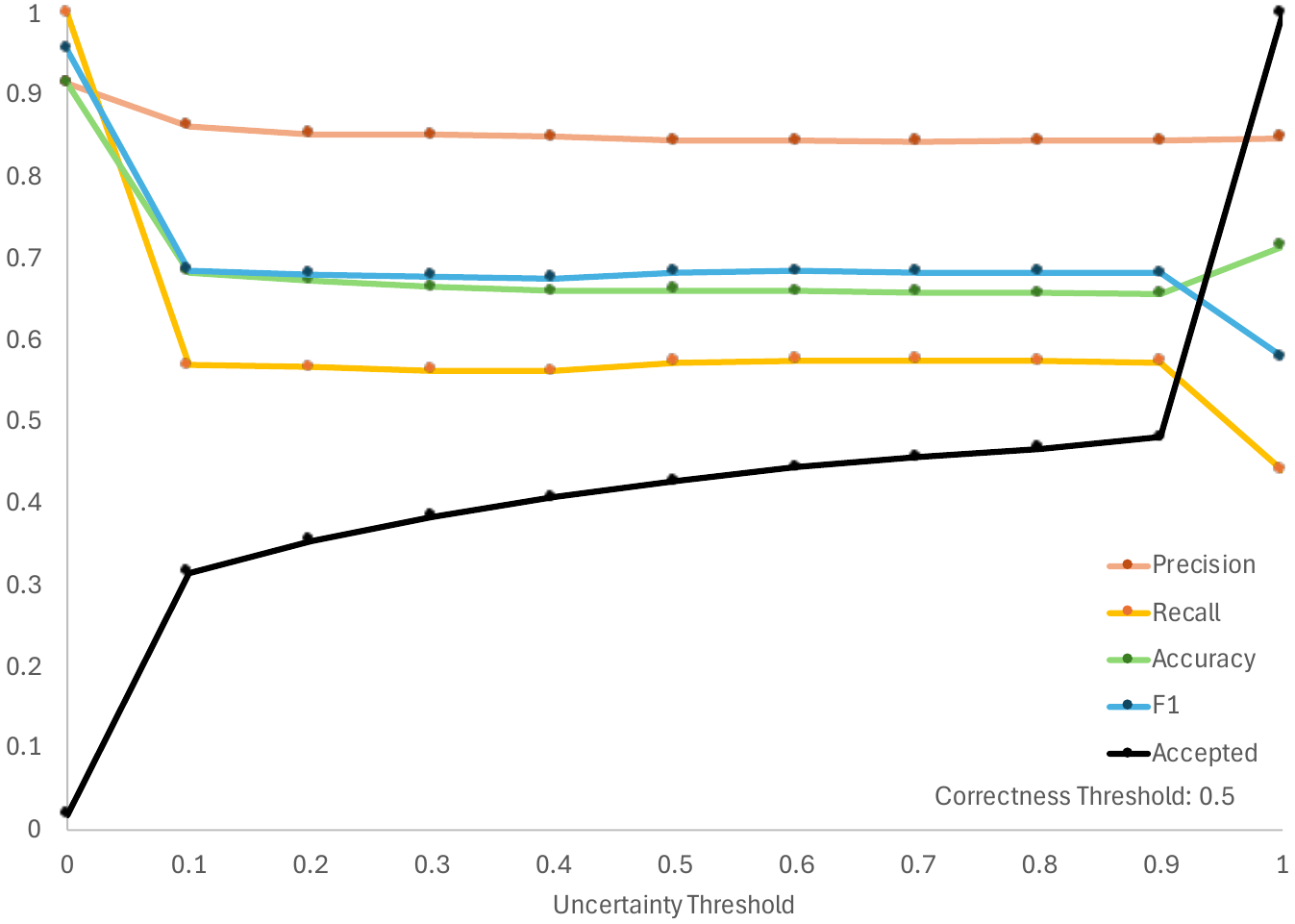}
\includegraphics[width=0.49\textwidth]{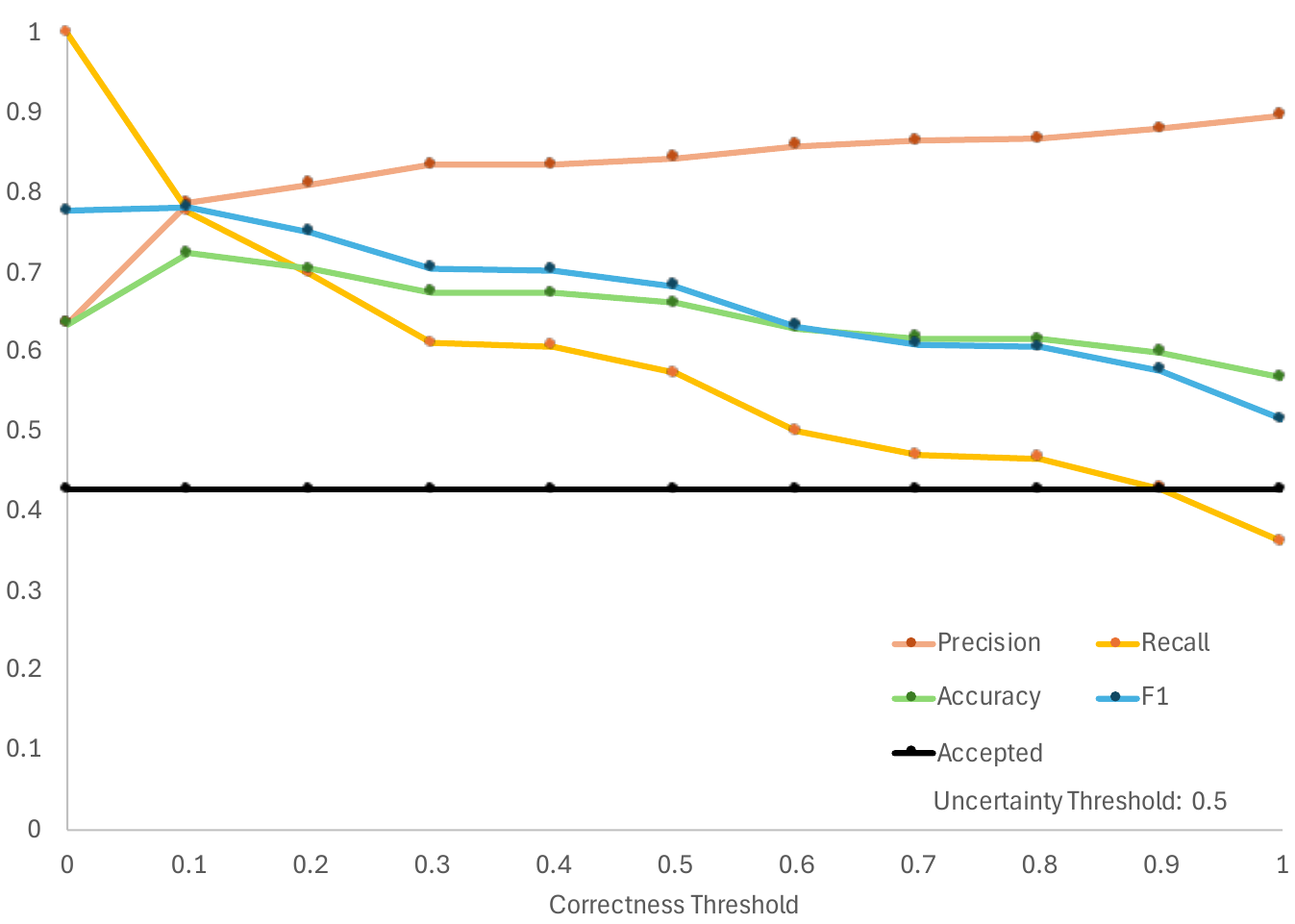}

\includegraphics[width=0.49\textwidth]{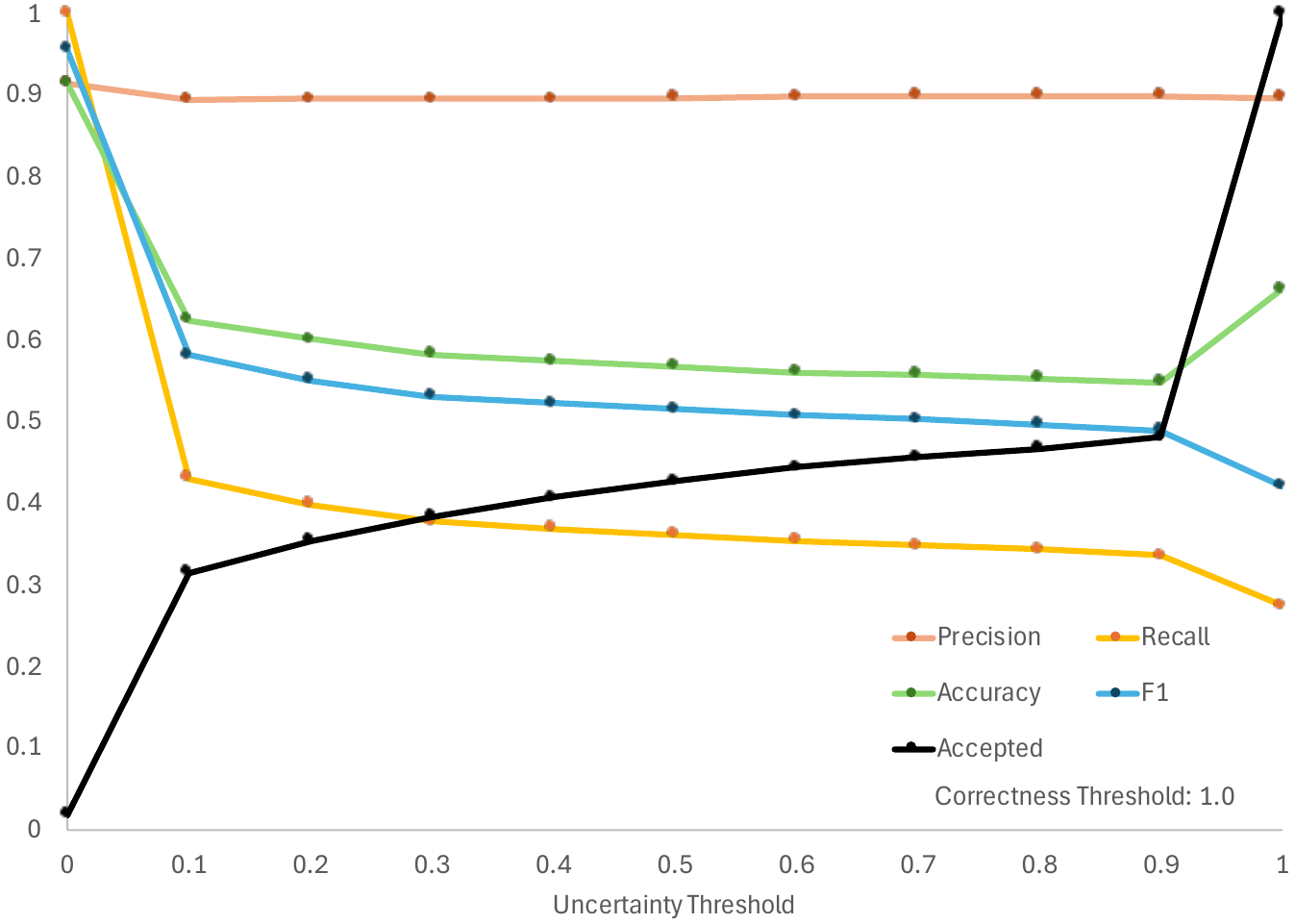}
\includegraphics[width=0.49\textwidth]{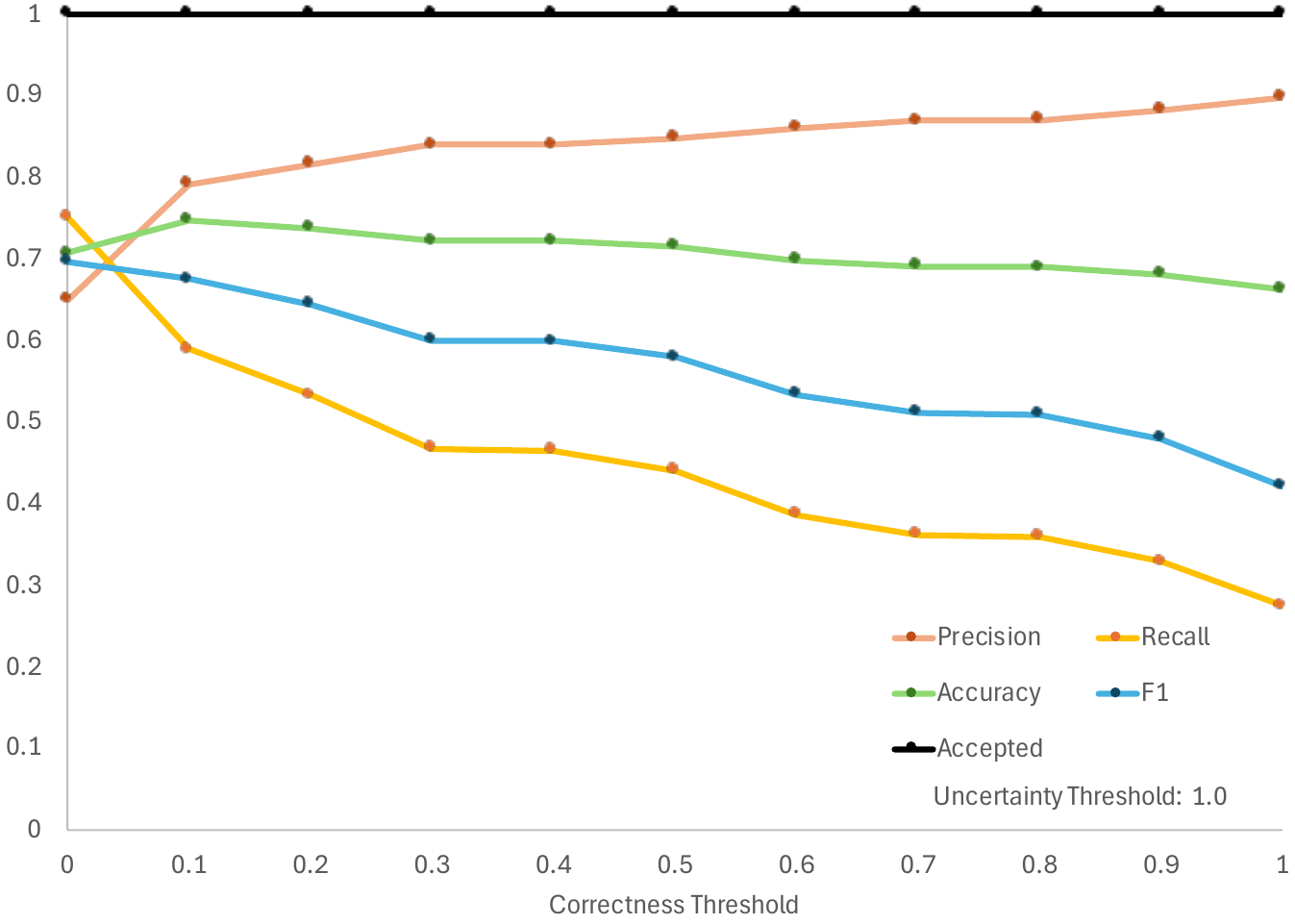}
\end{center}
\caption{Performance of the model for different values of minimum correctness and maximum uncertainty.}
\label{fig:performance}
\end{figure*}

Figure~\ref{fig:performance} shows the performance of the model as a function of the correctness and uncertainty thresholds. The top left panel is for a correctness threshold of 0.0, which means a rubric item would be graded as correct if any partial credit is assigned; thus, recall is very high, but accuracy and precision are low. At the other extreme, where full partial credit is required (correctness threshold 1.0), the opposite happens: precision is high but recall is low. Dependence on the uncertainty threshold is low, with a wide plateau except for the extremes of zero uncertainty or accepting any AI-grading. The right panels indicate the performance for fixed uncertainty threshold and differing correctness thresholds.

In each of the panels, indicated in black is the acceptance rate; naturally, the lower the uncertainty threshold, the more AI gradings will be rejected and delegated to human grading. As expected, adjusting the parameters of the model is a balancing act between accuracy and work savings.

\begin{figure}
\begin{tabular}{p{0.05\columnwidth}p{0.95\columnwidth}}
\begin{turn}{90} 
AI-assigned points
\end{turn}&

\includegraphics[width=0.94\columnwidth]{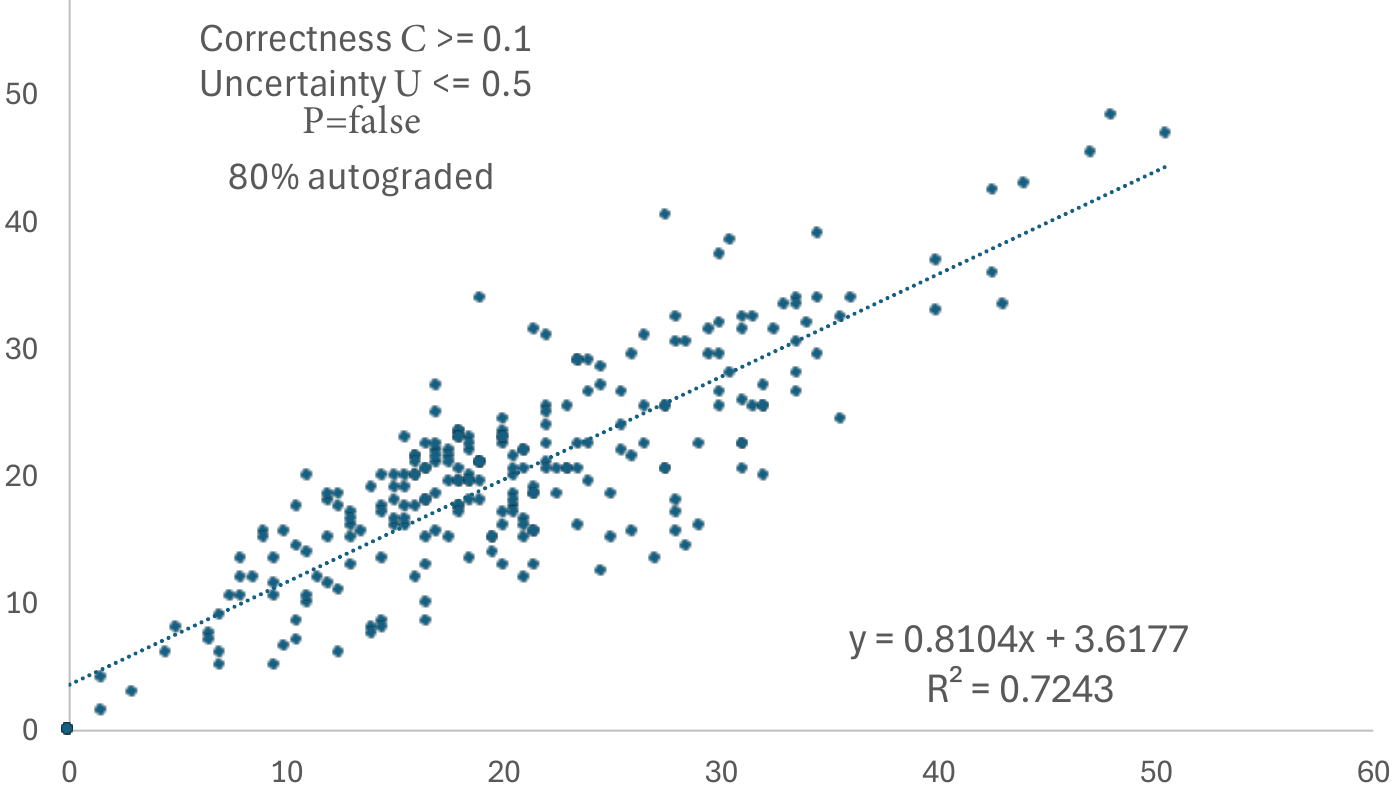}

\includegraphics[width=0.94\columnwidth]{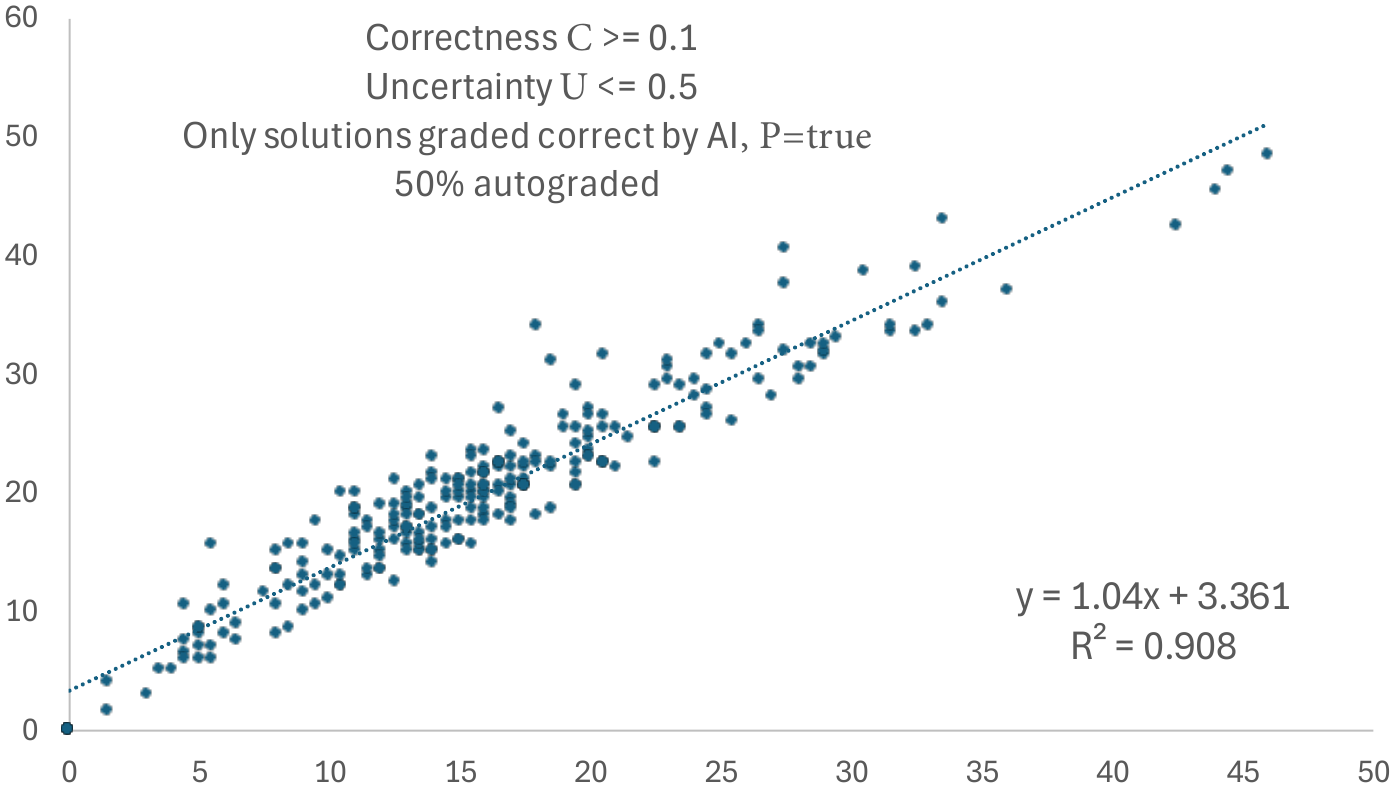}

\includegraphics[width=0.94\columnwidth]{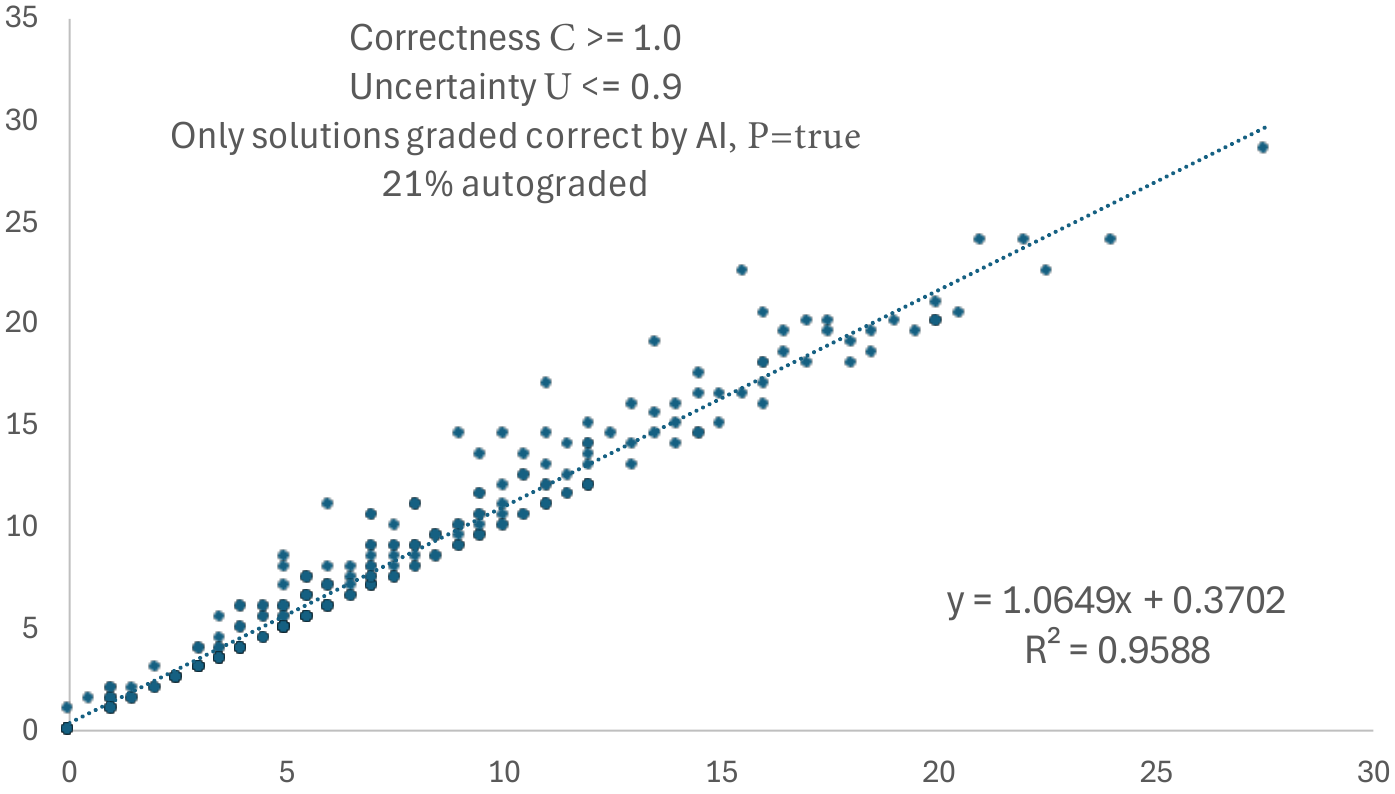}
\\
&
\begin{flushright}
TA-assigned points
\end{flushright}
\end{tabular}
\caption{Correlation between accepted AI-assigned points and the corresponding human-assigned points for the whole exam.}
\label{fig:corgraphs}
\end{figure}

Figure~\ref{fig:corgraphs} illustrates the effect of the correlation between accepted AI-assigned points and the corresponding human-assigned points for the whole exam. With $w_i$ being the point weight of rubric item $i$, and $t_{ij}$ being the points assigned by the TAs, we are thus comparing

\begin{equation}
\sum_{ij}\mathbf{1}_{\{A_{ij}\}}\mathbf{1}_{\{s_{ij}>C\}}w_i\quad\mbox{versus}\quad\sum_{ij}\mathbf{1}_{\{A_{ij}\}}t_{ij}
\end{equation}
Adding the filtering parameter $P$, for a correctness threshold of $C=0.1$ and an uncertainty threshold of $U=0.5$, the coefficient of determination $R^2$ increases from 0.72 to 0.91, while the fraction of automatically graded student-items decreases from 80\% to 50\%. The bottom panel shows a configuration where $R^2$ is 0.96 for about one fifth (21\%) of the student-item, meaning four-fifths of the items would need to be graded by humans. In both scenarios where only answers graded correct by the AI were accepted, the regression coefficient moved close to $1.0$, and the increase in correctness threshold brought the intercept nearer to $0.0$.
The highest achievable $R^2$ is 0.98, but in that case, only 3\% of the student-items are auto-graded.

\subsection{Outcome}
With respect to our hypotheses, we need to reject the hypothesis that the rubric items can be refined through statistical means. Effects were minimal, and mistakes were not caught.

The second hypothesis can be accepted. By using the three parameters --- minimum correctness, maximum uncertainty, and limiting acceptance to student-items graded as correct --- we were able to control grading accuracy in relation to the fraction of automated grading, with the parameters behaving as anticipated and producing consistent, explainable results.

\section{Discussion}
We were unable to iteratively improve the grading rubric using the employed statistical methods; in fact, while particularly the IRT-method turned out to be fickle, the overall grading performance was surprisingly robust against changes. This may be due to the fact that the problem text was included in the prompt, and the LLM was thus able to compensate even for typos. Based on our findings, there is no alternative to carefully proofreading the rubric items, but at the same time, not being overly concerned with details of their formulation.

In analyzing the impact of correctness and confidence thresholds on AI-assisted exam grading, it is crucial to determine which student-item combinations can reliably be assigned by the AI, and which should be subject to human review. The AI's grading performance is influenced by two adjustable factors: the correctness threshold and the uncertainty threshold. The correctness threshold determines how strictly the AI judges answers as correct based on a predefined score, while the uncertainty threshold filters predictions by the AI's confidence level. Lower uncertainty thresholds ensure that only predictions with high confidence are accepted, improving precision but reducing the number of predictions the AI makes. Conversely, higher uncertainty thresholds result in more predictions, though at the cost of lower precision, as the system begins to accept predictions with higher uncertainty.

The choice of correctness threshold also plays a critical role. Low thresholds allow the AI to predict more items as correct, increasing recall but potentially decreasing precision due to the inclusion of more false positives. In contrast, higher correctness thresholds improve precision by ensuring that only predictions with high AI scores are marked correct. However, this comes at the expense of recall, as fewer items are predicted as correct. Accuracy and the F1 score, which balances precision and recall, vary accordingly with changes in both thresholds. Thus, different threshold combinations affect both the workload on human graders and the reliability of the AI's predictions.

To optimize the balance between AI automation and human oversight, one must consider the trade-off between precision and recall. High precision thresholds are preferable when the goal is to minimize false positives, meaning the AI should only make predictions when it is highly confident of correctness. However, this approach leads to a higher workload for human graders since fewer predictions are accepted. On the other hand, prioritizing recall by lowering the correctness threshold allows the AI to identify more actual positives but increases the likelihood of errors, potentially requiring more post-processing or human intervention. The uncertainty threshold adds an additional layer of control, allowing predictions to be filtered based on the AI's confidence, thus directly influencing the number of accepted predictions and the need for human review.

In practical application, selecting optimal thresholds and selection criteria depends on the grading context. In high-stakes situations, precision should be prioritized to ensure that AI predictions are accurate and trusted. For lower-stakes exams, thresholds that strike a balance between precision and recall may be acceptable, allowing the AI to handle a larger portion of the grading load while maintaining a reasonable level of accuracy. Ongoing monitoring and adjustments of the thresholds based on empirical performance data will ensure that the AI system continues to support human graders effectively, maximizing both the reliability and efficiency of the grading process. Accepting only student-items considered correct had a strong influence on outcomes; in high-stakes situations, this is likely the best choice, while still providing work savings; as our results demonstrated, if the AI grades an item as correct, with very high probability it is indeed correct, while the opposite is not quite so certain.

Another practical concern is cost. We had each problem part graded ten times, but we would probably have been able to get away with less runs. The cost of \$7 per exam is comparable to the cost of one TA, assuming he or she gets paid \$20 per hour and spends about 20 minutes on grading one exam.

\section{Limitations}
This study is decidedly explorative and empirical, investigating different workflows for AI-assisted grading of one thermodynamics using models and tools currently available (Spring 2024). The results are specific to the exam, the rubric, the model, and the prompts, so only limited generalizability can be claimed. Students had to opt-in to the study, so self-selection bias cannot be excluded.

The study had a large number of participants and a large number of rubric items, which is a requirement for IRT to work reliably. The mechanism will break down if there are more rubric items than test takers.

\section{Outlook}
We plan to discontinue the investigation of the iterative rubric-refinement process in favor of more straightforward ``red flag'' mechanisms that detect clearly problematic rubric items. Rubrics should instead be carefully crafted and reviewed by human experts from the outset. In addition to including the problem statement in the system prompt (as shown here), grading accuracy may be gained by including course materials via Retrieval Augmented Generation (RAG)~\cite{lewis2020retrieval}, which is a technique we have already shown to be effective in customizing chatbots~\cite{kortemeyer2024ethel}. In particular, using RAG allows the grading step access to the course-specific notations and definitions.

The predictive mechanism for uncertainty determination explored in this study requires further validation using diverse datasets. Although the overall cost of \$7 per exam is comparable to the cost of employing TAs, conducting 10 independent grading runs for each student-item may have been excessive. Anecdotal evidence from grading logs suggests that for certain student responses, as few as three runs were sufficient to reach a final judgment --- particularly for responses that were either clearly correct or clearly incorrect, with no additional insights gained from the remaining seven runs. Moving forward, we aim to explore inter-rater reliability measures, such as Kappa statistics~\cite{hallgren2012computing}, during the grading process (as opposed to the simpler standard-deviation approach used here). These reliability measures could inform decisions on whether additional grading runs are necessary.

We have found that using GPT-4o is more accurate in transcribing handwriting than our previous approach with MathPix and GPT-4V~\cite{kortemeyer2024grading}, however, we found regional differences such as the handwriting of the digits 1 and 4 to be problematic. Additional research needs to go into the handwriting recognition of mathematical expressions; this step is crucial, as experiments with synthetic datasets generated directly  in LaTeX have demonstrated the crucial influence of the OCR step~\cite{kortemeyer24aigrading}.
 
To scale these mechanisms beyond research settings, considerable software engineering efforts will be required for efficient workflow management. Once exams have been scanned, the paper copies can be securely stored, with the remaining workflow conducted electronically. Student responses flagged by the uncertainty mechanisms should be routed to human graders, with an efficient load distribution system in place to ensure grading tasks are equitably assigned. Further into the future, the AI-grading may learn from human grader corrections in real-time, improving future grading decisions without requiring manual reprogramming of rubric rules.

\section{Conclusion}
This study explored the use of AI-assisted grading in high-stakes physics exams, focusing on how psychometric methods, particularly Item Response Theory (IRT), could enhance grading validity and support the iterative refinement of grading rubrics. While AI demonstrated potential in automating parts of the grading process, our findings underscore the complexities of integrating AI systems with human oversight in a way that ensures both accuracy and fairness.

Our results show that while psychometric approaches like IRT help quantify grading performance, they are limited in detecting or correcting rubric issues. Despite several iterations of rubric refinement based on statistical analysis, the overall impact on grading accuracy was minimal. This suggests that while AI can effectively grade with well-structured rubrics, automated refinement of these rubrics remains elusive. Human proofreading and careful rubric design are still critical, particularly in the context of complex problem-solving processes, where unexpected student approaches may challenge AI's capacity to adhere strictly to rubric guidelines.

The second aspect of our study, evaluating AI's grading reliability through correctness and uncertainty thresholds, yielded more promising results. By adjusting these thresholds, we demonstrated that AI can achieve high precision, particularly when focusing on items it grades as correct with high confidence. This flexibility allows instructors to balance grading accuracy with workload savings, delegating routine tasks to AI while retaining human review where needed. In high-stakes assessments, where minimizing false positives is essential, stricter thresholds may ensure greater trust in AI's performance, albeit with increased human intervention.

In practical terms, the financial cost of AI grading is comparable to that of human graders, suggesting that AI could be a cost-effective supplement to traditional grading, especially in large-scale assessments. However, optimizing this process requires careful consideration of how many independent AI grading runs are necessary to ensure reliable results.

In conclusion, while AI shows promise in supporting grading, its role is best seen as an aid rather than a replacement for human judgment. AI can reduce the grading burden for instructors by accurately handling routine tasks, but human oversight remains indispensable, particularly in high-stakes assessments where fairness and precision are paramount. With further refinement, AI-assisted grading has the potential to become a reliable and cost-effective tool in education, provided it is implemented with careful attention to grading validity and psychometric principles.

\begin{acknowledgments}
We thank David Pritchard for inspiring discussions at PERC 2024.
We would also like to thank the students who participated in this study, as well as Daria Onishchuk and Alina Yaroshchuk who assisted with logistics and the data analysis that laid the foundations for this study. We also thank Christine Kortemeyer for proofreading and commenting on the manuscript. This study is part of project Ethel~\cite{kortemeyer2024ethel}.
\end{acknowledgments}

\bibliography{GPT_IRT_exam}

\end{document}